\documentclass[article,showkeys,showpacs,preprintnumbers,amsmath,amssymb,aps]{revtex4-1}
\usepackage{graphicx}
\usepackage{dcolumn}
\usepackage{bm}
\begin{document}
\preprint{APS/123-QED}
\title{Conformal Invariance and  Warped 5-Dimensional Spacetimes}
\author{Reinoud Jan Slagter}
\email{info@asfyon.com}
\affiliation{ Asfyon, Astronomisch Fysisch Onderzoek Nederland,  1405EP Bussum \\ and \\ Department  of Physics, University of Amsterdam, The Netherlands}
\date{\today}
\begin{abstract}
We show that the Einstein field equations for a  five-dimensional warped spacetime, where only gravity can propagate into the bulk, determine the dynamical evolution of the warp factor of the four-dimensional brane spacetime. This can be explained as a holographic manifestation.
The warped 5D model can be reformulated by considering the warp factor as a dilaton field ($\omega$) conformally coupled to gravity and  embedded in a smooth $M_4 \otimes R$ manifold.
On the brane, where the U(1) scalar gauge fields live, the dilaton field manifests itself classically as a warp factor and enters  the evolution equations for the metric components and matter fields.
We write the Lagrangian for the Einstein-scalar gauge fields in a conformal invariant setting. However, as expected, the conformal invariance is broken (trace-anomaly) by the appearance of a mass term and a quadratic term in the energy-momentum tensor of the scalar gauge field, arising from the extrinsic curvature terms in the projected Einstein tensor.
These terms can be interpreted as a constraint in order to maintain conformal invariance.
By considering the dilaton field and Higgs field on equal footing on small scales, there will be no singular behavior, when $\omega\rightarrow 0$ and one can deduce constraints to maintain regularity of the action.
Our conjecture is that $\omega$, alias warp factor, has a dual meaning. At very early times, when $\omega \rightarrow 0$, it describes the small-distance limit,  while at later times it is a warp (or scale) factor that determines the dynamical evolution of the universe.
We also present a numerical solution of the model and calculate the (time-dependent) trace-anomaly. The solution depends on the mass ratio of the scalar  and gauge fields, the parameters of the model and the vortex charge $n$.
\end{abstract}
\pacs{11.27.+d, 11.10.Lm, 11.15.Wx, 12.10.-g, 11.25.-w, 04.50.-h, 04.50.Gh, 04.20.Ha}
\keywords{Conformal Invariance, Brane world models, U(1) scalar-gauge field, Cosmic strings, Dilaton field}
\maketitle

\section{Introduction}
The easiest way to modify general relativity (GR) is to extend the spacetime to more than four dimensions. Modification seems to be necessary in order to overcome the serious problems which one encounters when one decreases the scale closer to the Planck scale. Specifically, the problems enclose the hierarchy problem, the cosmological constant problem, the fate of the black hole, the issue of dark energy and last but not least the handling of scales.
There seems to be no limit on the smallness of fundamental units in one particular domain of physics, while in others there are very large space and time scales.

A very attractive higher dimensional model is the so-called warped spacetime of Randall and Sundrum (RS)\cite{ran1}. In this model one assumes that there is one large extra dimension.
The result is that  effective 4D Kaluza-Klein(KK) modes are obtained from the perturbative  5D graviton. These KK modes will be massive from the brane viewpoint.
The modified  Einstein equations on the brane  and scalar gauge field equations will now contain contributions from the 5D Weyl tensor\cite{roy1,roy2,roy3,shir}. The hierarchy problem is solved in these models, because the graviton's probability function is extremely high at the Planck-brane and drops exponentially as it moves closer towards the TeV-brane.

Warped spacetimes can also be linked to conformal symmetry. Conformal invariance(CI) is an approved property in string theory by the Anti-DeSitter/Conformal Field Theory (AdS/CFT) correspondence, where a conformal field theory sits on the boundary of the Anti-DeSitter spacetime.
Some decades ago, 't Hooft\cite{thooft5}  proposed that the information about an extra dimension is visible as a curvature in a spacetime with one fewer dimension. So the fifth dimension can act as a spacetime fabric on the 4D boundary.
The appearance of the five-dimensional curved spacetime is also natural from the black hole entropy. A five-dimensional black hole has entropy which is proportional to the five-dimensional “area.” But an “area” in five dimensions is a “volume” in four dimensions: this is appropriate as the entropy of a four-dimensional statistical system.
Later it was realised that this principle can be reformulated as so-called AdS/CFT duality models.
A famous example is the type IIB-string theory on the background of $AdS_5\otimes S^5$. It is equivalent to a (3+1)D super Yang-Mills model with U(N) symmetry living on the boundary.
The AdS/CFT correspondence could even solve the black hole information paradox ( i.e., the unitarity paradox of time evolution\cite{thooft2,thooft3}).
In the model of RS, the gravitational degrees of freedom of the extra dimension appear on the brane as a dual field theory under AdS/CFT correspondence. So holography could be a prerequisite for the existence of such models.
In conformal GR with (quantum) fields, the dilaton plays a fundamental role. In the low-energy limit the dilaton field can act as  a dynamical warp factor in 5D warped spacetimes.
Because there is strong evidence  that our universe is now expanding at an accelerating rate, a desirable situation would be that at earlier  times the AdS/CFT correspondence holds and at later limes a dS/CFT correspondence\cite{strom}. The warp factor could contribute to such a model.

At any level, CI in GR remains a peculiar issue. There is the question  if it  is possible to incorporate other fields ( massless and massive) into CI GR\cite{thooft2,thooft3,thooft1}. The resulting model should explain why in the high-energy situation mass scales are unimportant and could be of help to construct singularity-free spacetime by pushing them to infinity.
Further, the notion of conformal null infinity and the definition of energy flux can be formulated.
Conformal gauge-fixing procedures can also be linked by the upper limit of the amount of information that can be stored in a 5D spacetime, i.e., on a 4D hypersurface\cite{thooft4}.
The model  has also shortcomings, described as anomalies. Will all the beta-functions of the conformal model vanish? It is hoped that at the quantum level anomalies can be removed and a kind of spontaneous symmetry breaking can be formulated at lower energies.
In any case, non-conformal mass terms in the Lagrangian (for example  a scalar gauge field with a potential term), does not affect the  CI of the effective action after integrating over $\omega$ (dilaton)\cite{thooft2}.
We also have  the problem of the cosmological constant. A dimensionful mass term in the potential of the Higgs field breaks the tracelessness of the energy momentum tensor. However, it turns out that the tracelessness of the energy momentum tensor can be maintained if a cosmological constant is also generated\cite{mann}.

A related problem is the asymptotic flatness of isolated systems in GR, specially when they radiate. There is a back-reaction of disturbances on the background metric: we have no flat metric in terms of which the falloff of the curvature can be specified. There are other problems which can be linked to CI, i.e., the notion of asymptotic flatness at null infinity, topological  regularity,  the gravitational energy emitted by compact objects  and how to handle the limit as one approaches  infinity (see for example the textbook of Wald\cite{wald}). Further, one needs a strongly asymptotically predictable spacetime and  the redefinition $ g_{\mu\nu}=\omega^2 \tilde g_{\mu\nu}$. Here $\omega$ represents the dilaton field, or conformal factor and must be handled on an equal footing as the Higgs field. The "un-physical" metric $\tilde g_{\mu\nu}$ must then be generated from at least Ricci-flat spacetimes.
However, one should like to have $\tilde g_{\mu\nu}=\eta_{\mu\nu}$, the flat Minkowski, close to the Planck scale. The challenge is therefore to investigate the possibility that $\tilde g_{\mu\nu}$ is emergent during the evolution of our universe.
In general, one could say that a conformal structure for gravity is inevitable and is the missing symmetry for spacetimes.

In this manuscript we reformulate the results found earlier\cite{slag1} in the light of conformal invariance. In this former model, a warped U(1) scalar gauge field could also be used to explain the curious alignment of the polarization axes of quasars in large quasar groups on Mpc scales\cite{slag2,slag3}. This is possible, because a profound contribution to the energy-momentum tensor comes from the bulk spacetime and can be understand as "dark"-energy. The scalar field becomes super-massive by the contribution of the 5D Weyl tensor on the brane and stored azimuthal preferences of the spinning axes of the quasars just after the symmetry breaking.
The outline of this manuscript is as follows. In section 2 we discuss the 5D warped spacetime. In section 3 we reformulated the model in a conformal way. In sections 4 and 5 we handle the "un-physical" metric without matter and in section 6 we add matter to the model and discuss the breaking of the conformal invariance.
\section{The warped 5D spacetime with a U(1) scalar gauge field}
\label{sec:2}
Let us consider  a warped five-dimensional FLRW spacetime\cite{slag1}
\begin{eqnarray}
ds^2 = {\cal W}(t,r,y)^2\Bigl[e^{2\gamma(t,r)-2\psi(t,r)}(-dt^2+ dr^2)+e^{2\psi(t,r)}dz^2+\frac{r^2}{e^{2\psi(t,r)}}d\varphi^2\Bigr]+ \Gamma(y)^2dy^2,\label{eqn1}
\end{eqnarray}
with ${\cal W}=W_1(t,r)W_2(y)$ the warp factor. Our 4-dimensional brane is located at $y=0$. All standard model fields reside on the brane, while gravity can propagate into the bulk.
The 5D Einstein equations are
\begin{equation}
{^{(5)}G}_{\mu\nu}=-\Lambda_5{^{(5)}g}_{\mu\nu}+\kappa_5^2 \delta(y)\Bigl(-\Lambda_4 {^{(4)}g}_{\mu\nu}+{^{(4)}T}_{\mu\nu}\Bigr), \label{eqn2}
\end{equation}
with $\kappa_5= 8\pi {^{(5)}G}= 8\pi/{^{(5)}M}_{pl}^3$, $\Lambda_4$ the brane tension, ${^{(4)}g}_{\mu\nu}={^{(5)}g}_{\mu\nu}-n_\mu n_\nu$ and $n^\mu$ the unit normal to the brane. The ${^{(5)}M}_{pl}$ is the fundamental 5D Planck mass, which is much smaller than the effective Planck mass on the brane, $\sim 10^{19}$ GeV. We consider here the matter field ${^{(4)}T}_{\mu\nu}$ confined to the brane, i.e., the U(1) scalar gauge field.

From the combination of the components of the 5D Einstein equations, ${{^5}G}_{tt}-{{^5}G}_{rr}$ one obtains for $W_1(t,r)$ the partial differential equation\cite{slag1}
\begin{equation}
\partial_{tt}W_1=\partial_{rr}W_1+\frac{1}{W_1}\Bigl((\partial_r W_1)^2-(\partial_t W_1)^2\Bigr)+\frac{2}{r}\partial_r W_1. \label{eqn3}
\end{equation}
A typical solution is
\begin{equation}
W_1(t,r)=\frac{\pm1}{\sqrt{\tau r}} \sqrt{\Bigl(d_1 e^{(\sqrt{2\tau})t}-d_2e^{-(\sqrt{2\tau})t}\Bigr)\Bigl(d_3 e^{(\sqrt{2\tau})r}-d_4e^{-(\sqrt{2\tau})r}\Bigr)}, \label{eqn4}
\end{equation}
with $\tau, d_i$ some constants. So we have two branches, i.e., the plus and minus sign in Eq. (\ref{eqn4}). $W_1$ can also be complex\footnote{ Eq. (\ref{eqn3}) is invariant under $W_1\rightarrow i W_1$}. In figure 1 we plotted typical solutions of $W_1$.
The dynamics of the "scale-factor" of the 4D hyper-surface is solely determined by the 5D Einstein equations (in the classical 4D situation one easily obtains from the Einstein equations that  $W_1(r,t)$ must be $r-$independent).
The y-dependent equations
\begin{eqnarray}
\partial_{yy}W_2=-\frac{(\partial_y W_2)^2}{W_2}-\frac{1}{3}\Lambda_5W_2\Gamma^2-\frac{c_1}{W_2}+\frac{\partial_y W_2\partial_y\Gamma}{\Gamma}, \quad
(\partial_y W_2)^2=-\frac{1}{6}\Lambda_5W_2^2\Gamma^2+c_2\Gamma^2, \label{eqn5}
\end{eqnarray}
yield the well-known solution (for $\Gamma(y)=1$)
\begin{equation}
W_2(y)=e^{\sqrt{- \frac{1}{6} \Lambda_5}(y- y_0)} \label{eqn6}
\end{equation}
of the Randall-Sundrum model\cite{ran1}.
It is remarkable that the function $\Gamma(y)$ is undetermined by the 5D field equations. So the initially non-factorizable geometry of Eq.(\ref{eqn1}) can in fact be written as $M_4 \times K$, with $K$ an Euclidean smooth compact manifold.
The effective 4D Einstein-Higgs-gauge field equations on the brane\cite{shir}
\begin{eqnarray}
{^{4}\!G}_{\mu\nu}=-\Lambda_{eff}{^{4}\!g}_{\mu\nu}+\kappa_4^2 {^{4}\!T}_{\mu\nu}+\kappa_5^4{\cal S}_{\mu\nu}-{\cal E}_{\mu\nu}, \label{eqn7}
\end{eqnarray}
contain a contribution  ${\cal E}_{\mu\nu}$ from the 5D Weyl tensor and carries information of the gravitational field outside the brane. The quadratic term in the energy-momentum tensor, ${\cal S}_{\mu\nu}$, arises from the extrinsic curvature terms in the projected Einstein tensor. ${^{4}\!T}_{\mu\nu}$ represents the matter content on the brane, in our case, the scalar and gauge fields $\Phi=\eta X(t, r)e^{i n\varphi}, A_\mu =\frac{n}{ \epsilon }\bigl(P(t, r)-1\bigr)\nabla_\mu\varphi$ and contains the potential term $V(\Phi)=\frac{1}{8}\beta(\Phi \Phi^*-\eta^2)^2$ in the case of the Higgs field. n represents the  multiplicity ( or winding number) of
the Higgs field\footnote{Sometimes one absorbs n in the gauge field by writing  $A_\mu =\frac{1}{\epsilon}(P-n)\partial_\mu\varphi$}.
Further, $\Lambda_{eff}=\frac{1}{2}(\Lambda_5+\frac{1}{6}\kappa_5^4\Lambda_4^2)$. In the low energy limit one recovers conventional Einstein gravity.
From the 4D effective Einstein equations one cannot isolate an  equation for $W_1$. So $W_1$ is a warp factor effect. The equations for the metric components $\gamma$ and $\psi$ together with the scalar and gauge field, can be solved numerically\cite{slag1}, with $W_1$ given by Eq.(\ref{eqn4}).
\begin{figure}[h]
\centerline{
\includegraphics[width=4.2cm]{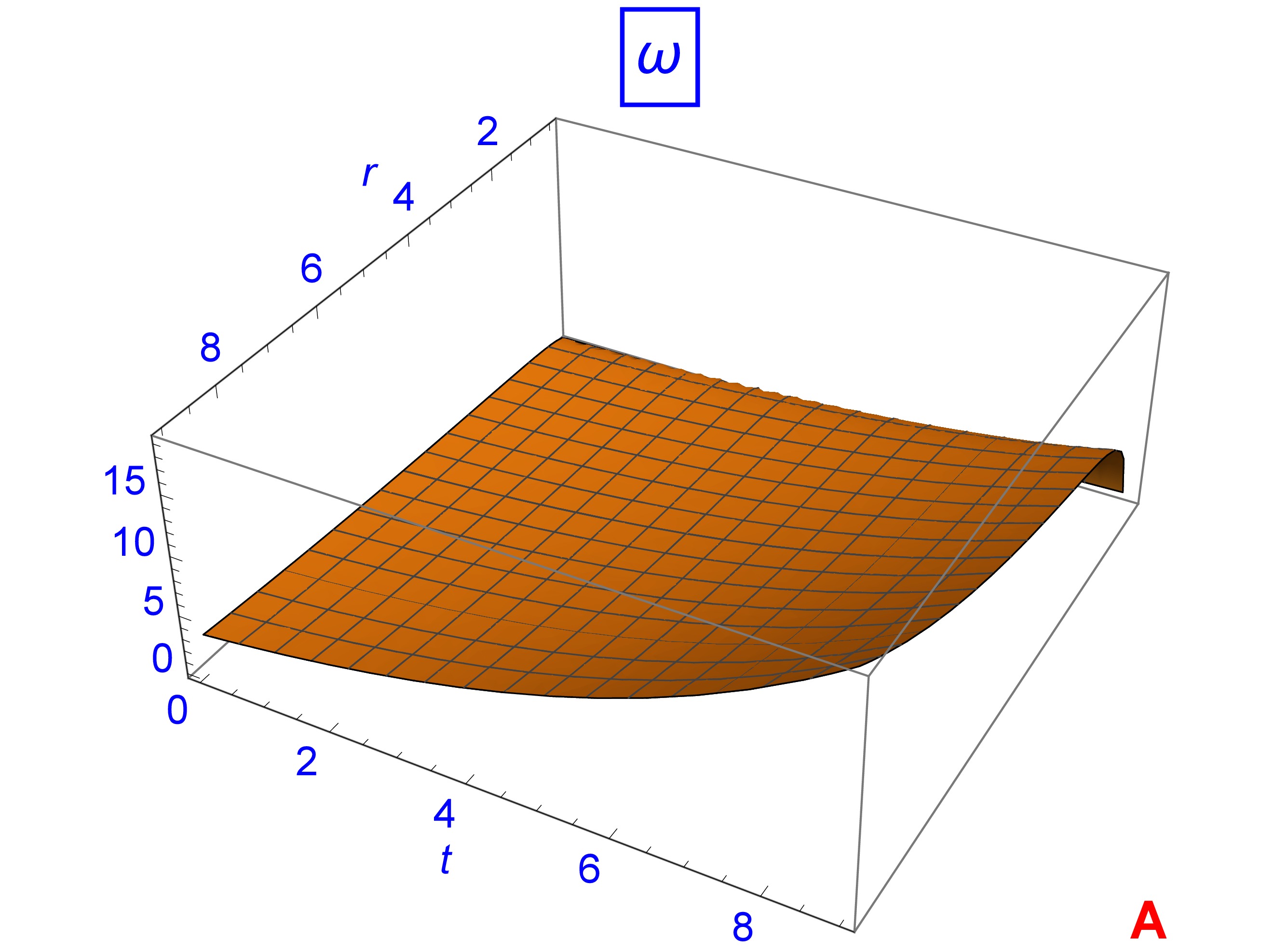}
\includegraphics[width=4.2cm]{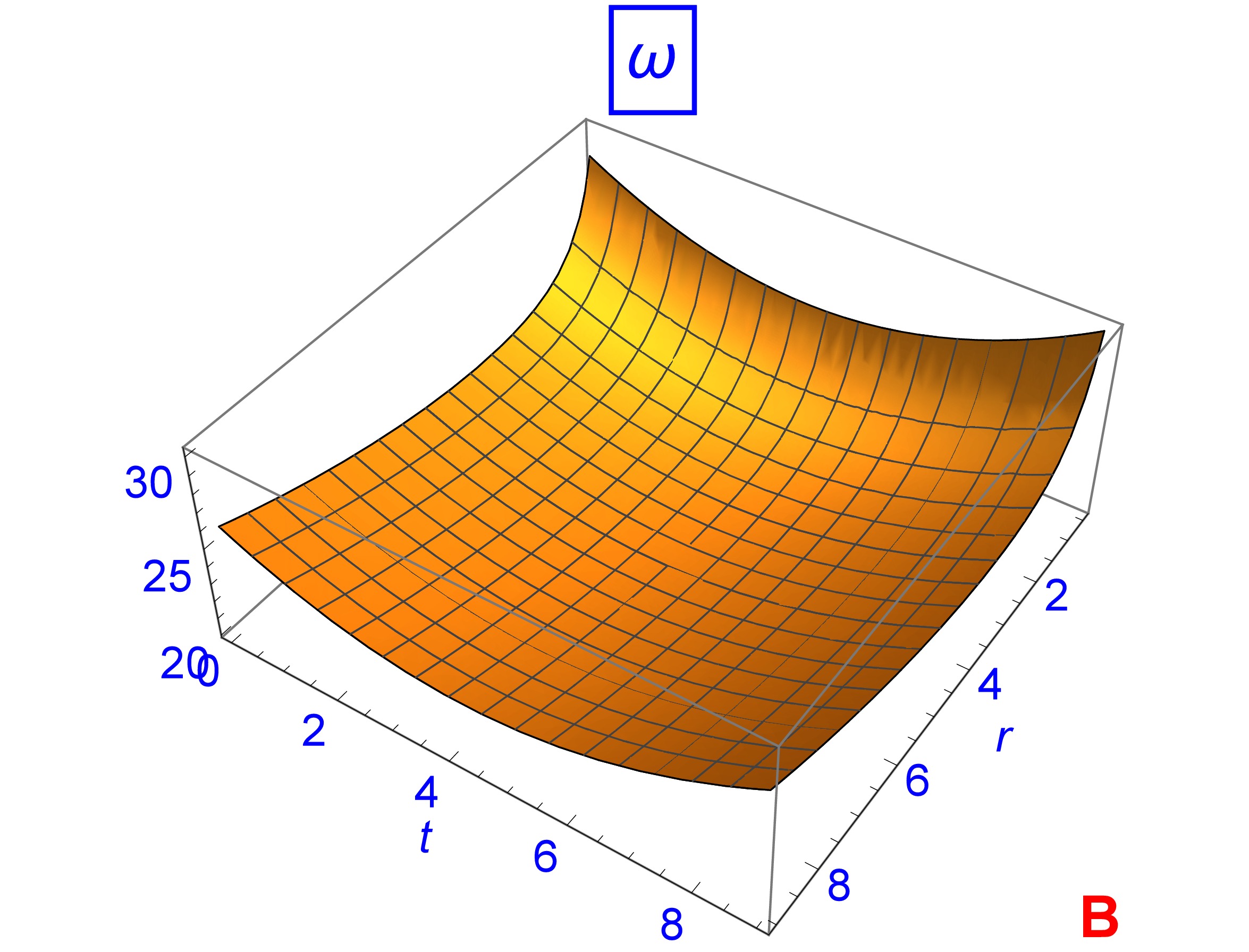}
\includegraphics[width=4.2cm]{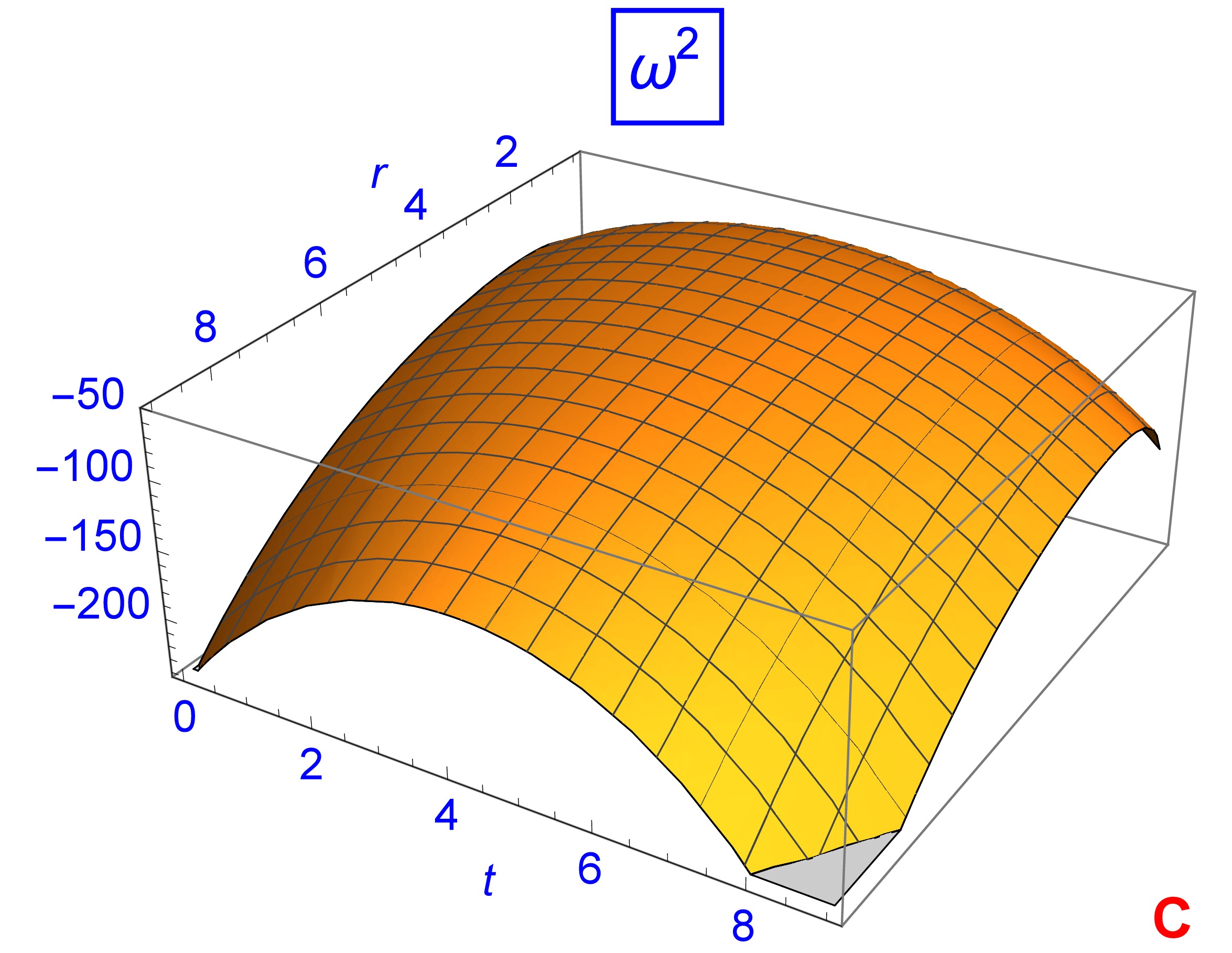}
}
\vskip 0.4cm
\caption{Three different plots of the warp factor $W_1$, reinterpreted as dilaton field $\omega$. Solution C is complex. See text. }
\end{figure}
It is of interest to observe that our "scale"-factor $W_1$ is $(t,r)$-dependent. This implies that the expansion of the universe is different at different moments in time.
The acceleration of our universe, with equation of state $p=w\rho$ with $w$ close to $-1$, can then  be explained without the need of a cosmological constant ( or dark energy). The Einstein equations and the scalar gauge field equations are modified by the presence of the warp factor.

\section{The warped 5D metric revisited}
\label{sec:3}
Let us return to our metric Eq.(\ref{eqn1}) and define
\begin{equation}
g_{\mu\nu}=\omega_1(t,r)^2 W_2(y)^2\tilde g_{\mu\nu}+n_\mu n_\nu \Gamma(y)^2.\label{eqn8}
\end{equation}
The $\tilde g_{\mu\nu}$ is sometimes called the un-physical metric. We now write the warp factor $W_1$ as a dilaton field $\omega_1$, which satisfy  Eq.(\ref{eqn3}), written as ( we omit the index 1)
\begin{equation}
\partial_{tt}\omega=\partial_{rr}\omega+\frac{1}{\omega}\Bigl((\partial_r\omega)^2-(\partial_t\omega)^2\Bigr)+\frac{2}{r}\partial_r\omega. \label{eqn9}
\end{equation}
This was only possible, because we could separate Eq.(\ref{eqn3}) from the 5D Einstein equations.
We rewrite the solution for the  dilaton as (i.e., Eq.(\ref{eqn4}) )
\begin{equation}
\omega^2=\frac{1}{\tau r}\Bigl(d_1 e^{(\sqrt{2\tau})t}-d_2e^{-(\sqrt{2\tau})t}\Bigr)\Bigl(d_3 e^{(\sqrt{2\tau})r}-d_4e^{-(\sqrt{2\tau})r}\Bigr). \label{eqn10}
\end{equation}

We can interpret the warp factor, which originally described the behavior of the expansion of our warped spacetime, as a "scaling field" or dilaton field $\omega$ in conformal invariant gravity.  For metrics with $det(\tilde g_{\mu\nu})=-1$, it determines the scales for rulers and clocks (see for example 't Hooft's treatment of this issue\cite{thooft4}).

In general, if one considers  a field ${\cal F}$ on a metric $g_{\mu\nu}$, one says that $\Omega^s{\cal F}$ is conformally invariant with metric $\Omega^2 g_{\mu\nu}$ for all conformal factors $\Omega^2$. s is called the conformal weight of the matter field.

We consider now the conformally invariant Lagrangian without matter
\begin{eqnarray}
{\cal L}^{EH\omega}=\frac{\sqrt{-\tilde g}}{16\pi G}\Bigl(\omega^2\tilde R+6\tilde g^{\mu\nu}\partial_\mu\omega\partial_\nu\omega\Bigr)\label{eqn11}.
\end{eqnarray}
One could also add a cosmological term $-2\Lambda_4\omega^4$ (see section 6). Variation with respect to $\tilde g_{\mu\nu}$ results in the Einstein equation
\begin{eqnarray}
\tilde G_{\mu\nu}=\frac{1}{\omega^2}\Bigl[\tilde\nabla_\mu\tilde\nabla_\nu\omega^2-\tilde g_{\mu\nu}\tilde\nabla_\alpha\tilde\nabla^\alpha\omega^2-6(\partial_\mu\omega\partial_\nu\omega-\frac{1}{2}\tilde g_{\mu\nu}\partial_\alpha\omega\partial^\alpha\omega)\Bigr]\equiv \frac{1}{\omega^2}\tilde T_{\mu\nu}^{(\omega)}.\label{eqn12}
\end{eqnarray}
Variation of Eq.(\ref{eqn11}) with respect to $\omega$ yields the well-known conformal invariant equation
\begin{equation}
\tilde\nabla^\mu\partial_\mu\omega-\frac{1}{6}\omega \tilde R=0.\label{eqn13}
\end{equation}
One can easily verify that {\bf TR}$[\tilde G^{\mu\nu}-\frac{1}{\omega^2}\tilde T_{\mu\nu}^{(\omega)}]=0$ with the help of  Eq.(\ref{eqn13}). So the trace of any matter field contribution must be zero. We will return to this issue in section 6. Here we already remark that Maxwell's equations, $\nabla^\mu F_{\mu\nu}$, are conformal invariant (for dimension n=4), but  Laplace's equation for a scalar field $\Phi$, $\nabla^\mu\partial_\mu \Phi =0$, is not. One can easily proof that for n dimensions,
\begin{equation}
\nabla^\mu\partial_\mu\Phi-\frac{n-2}{4(n-1)}\Phi R=0,\label{eqn14}
\end{equation}
is conformal invariant for conformal weight $\frac{2-n}{2}$. This scalar-field equation follows also directly from Euler-Lagrange equations for the action\cite{wald}
\begin{equation}
{\cal I}=\frac{1}{16\pi G}\int d^nx\sqrt{-g}\Bigl[\Phi^2 R+4\frac{n-1}{n-2}g^{\mu\nu}\nabla_\mu\Phi\nabla_\nu\Phi\Bigr].\label{eqn15}
\end{equation}

We still can perform an additional gauge freedom, i.e., a local conformal transformation
\begin{eqnarray}
\tilde g_{\mu\nu}\rightarrow \Omega^2\tilde g_{\mu\nu}, \qquad \omega\rightarrow\frac{1}{\Omega}\omega, \qquad \Phi\rightarrow\frac{1}{\Omega}\Phi. \label{eqn16}
\end{eqnarray}
The transformation properties of $G_{\mu\nu}$ and  $R$ are
\begin{equation}
G_{\mu\nu}\rightarrow G_{\mu\nu}+\frac{2}{\Omega^2}\Bigl(2\tilde\nabla_\mu\Omega\tilde\nabla_\nu\Omega-\Omega\tilde\nabla_\mu\tilde\nabla_\nu\Omega-\frac{1}{2} \tilde g_{\mu\nu}(\tilde\nabla^\mu\Omega\tilde\nabla_\mu\Omega -2\Omega\tilde\nabla^\mu\tilde\nabla_\mu\Omega\Bigr),\label{eqn17}
\end{equation}
\begin{equation}
R\rightarrow\frac{1}{\Omega^2}\Bigl(R-\frac{6}{\Omega}\tilde\nabla^\mu\tilde\nabla_\mu\Omega\Bigr)\label{eqn18},
\end{equation}
so the vacuum breaks local conformal invariance. We obtain that $\Omega$  obeys the Laplace equation $\tilde\nabla^\mu\tilde\nabla_\mu \Omega=0$ as gauge condition if the Ricci scalar remains zero.
\section{The un-physical metric $\tilde g_{\mu\nu}$}
\label{sec:4}
We can solve the Einstein equations for $\tilde g_{\mu\nu}$ together with the dilaton solution and without  any matter.
From Eq.(\ref{eqn12})we obtain
\begin{equation}
\tilde G_{\mu\nu}=\frac{1}{\omega^2}\tilde T_{\mu\nu}^{(\omega)}-{\cal E}_{\mu\nu}.\label{eqn19}
\end{equation}
The extra term ${\cal E}$ comes from the projected Weyl tensor\cite{shir}, because we must use Eq.(\ref{eqn7}) as effective 4D Einstein equations.
\begin{figure}[h]
\centerline{
\includegraphics[width=5cm]{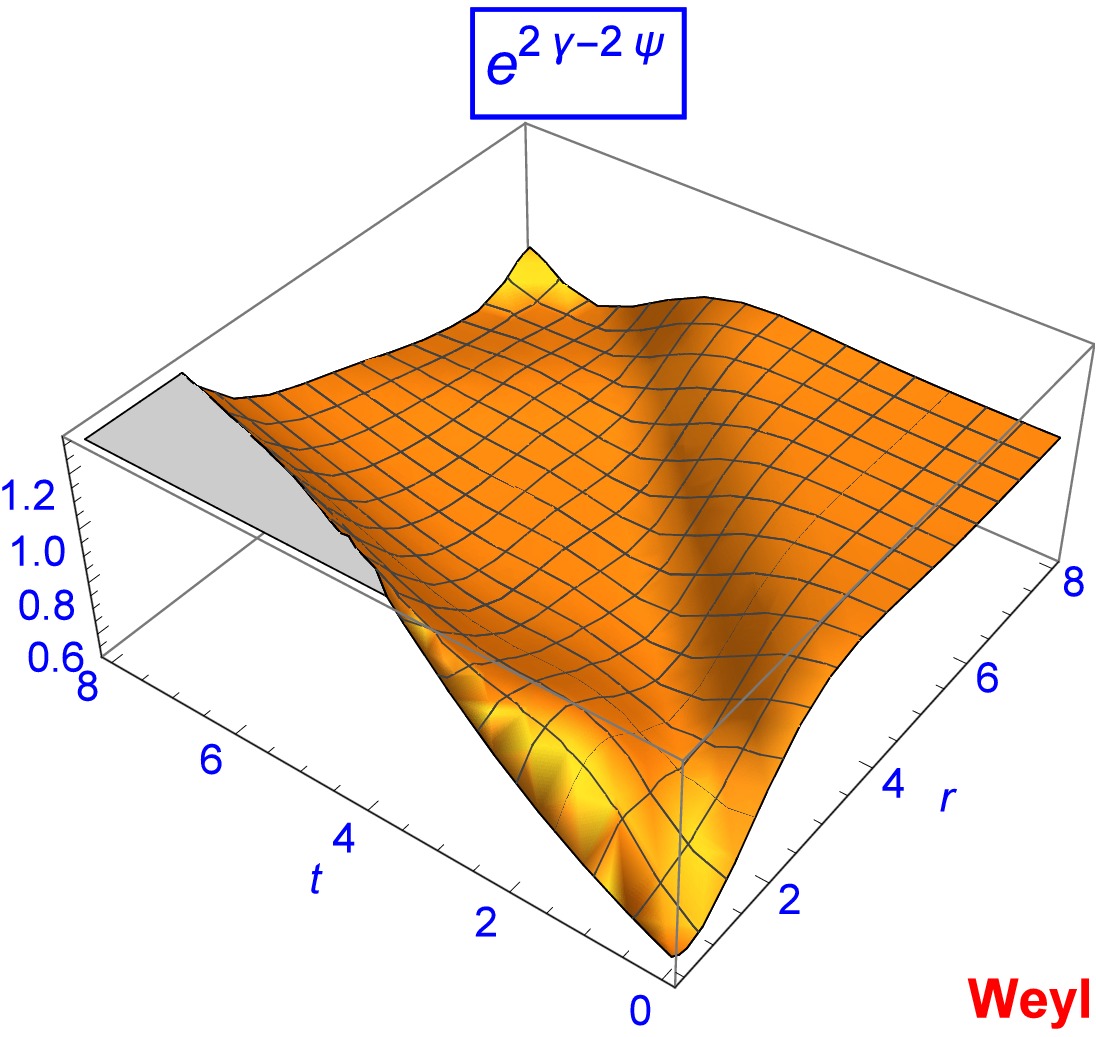}
\includegraphics[width=5cm]{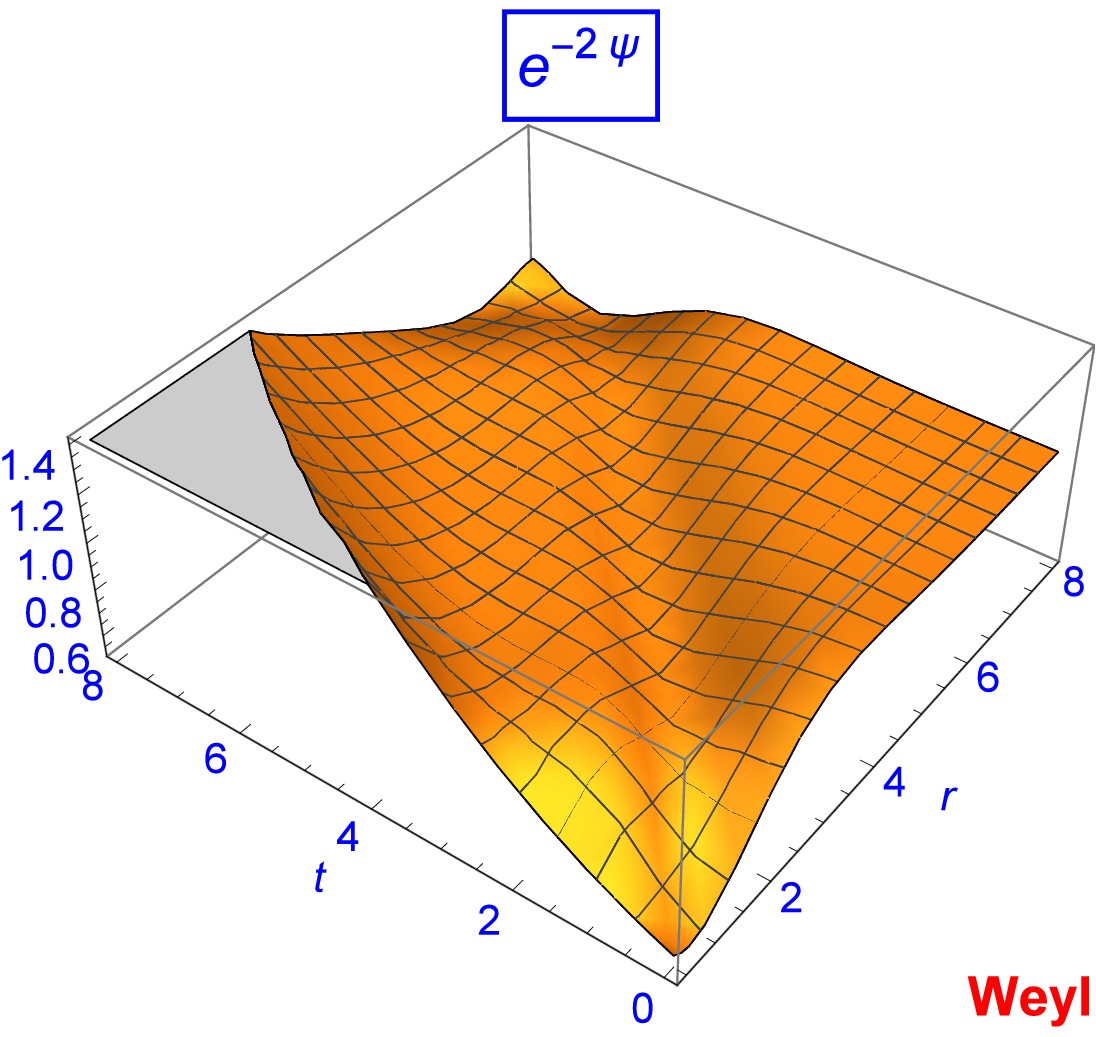}}
\centerline{
\includegraphics[width=5cm]{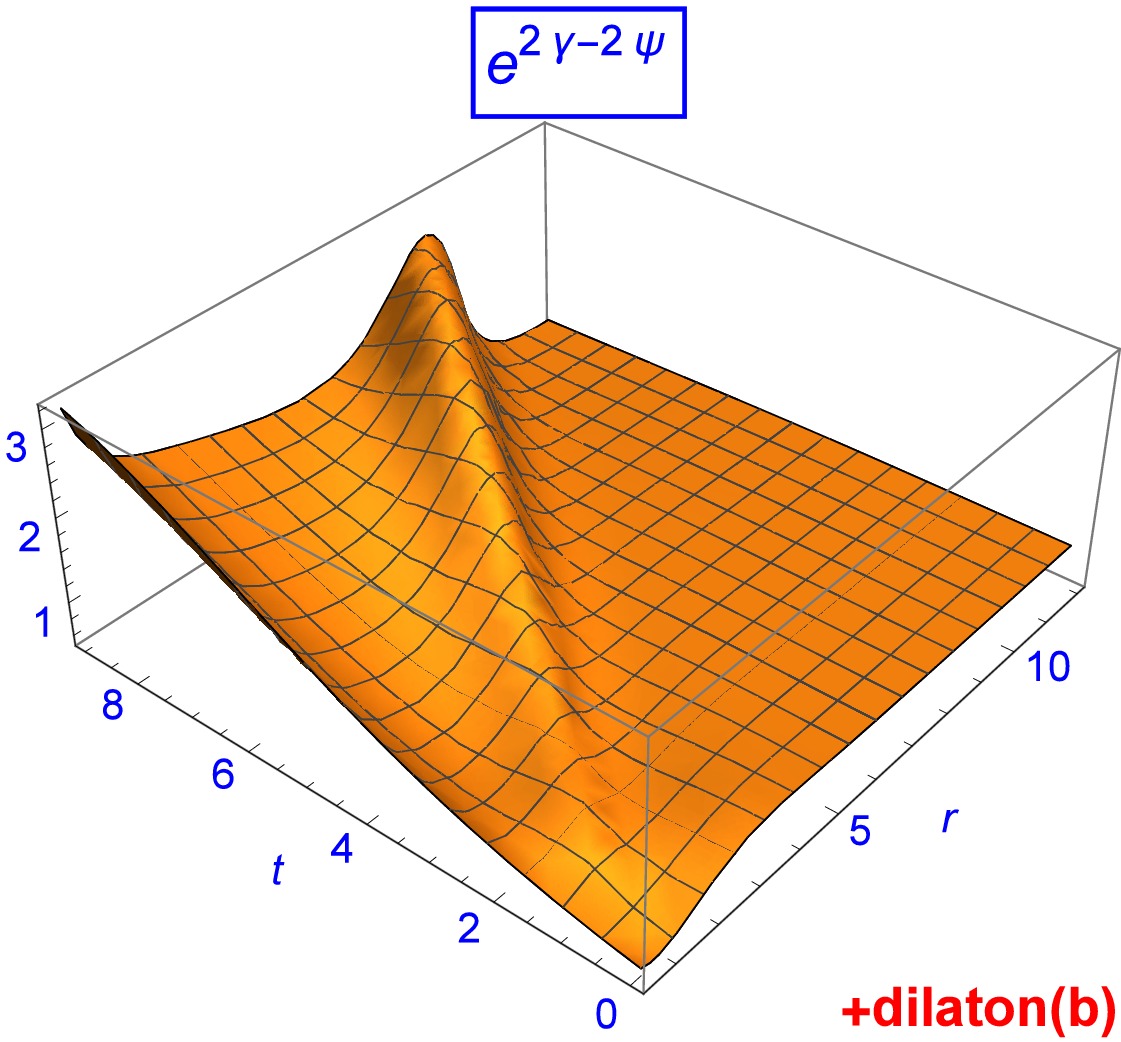}
\includegraphics[width=5cm]{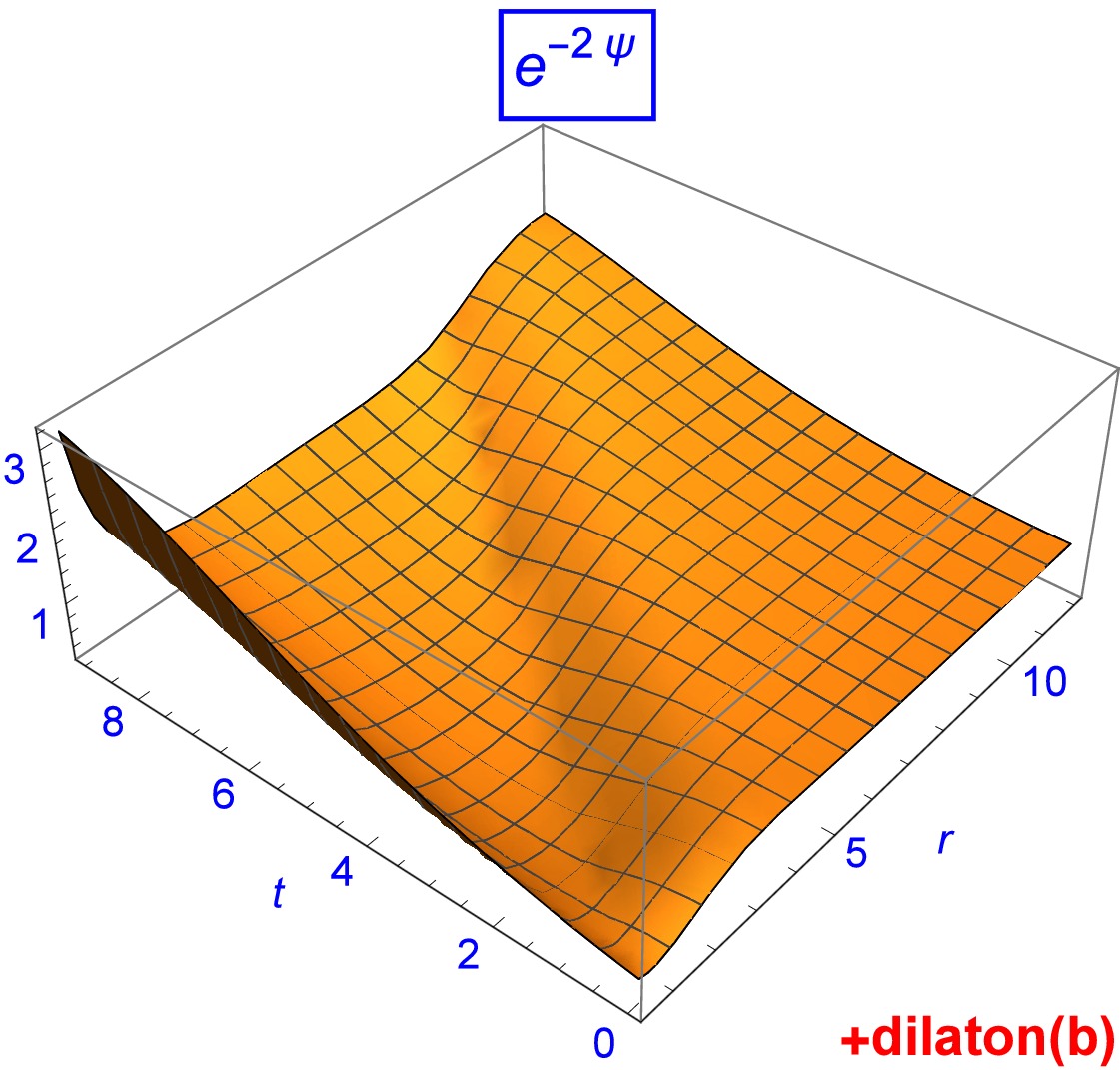}
\includegraphics[width=5cm]{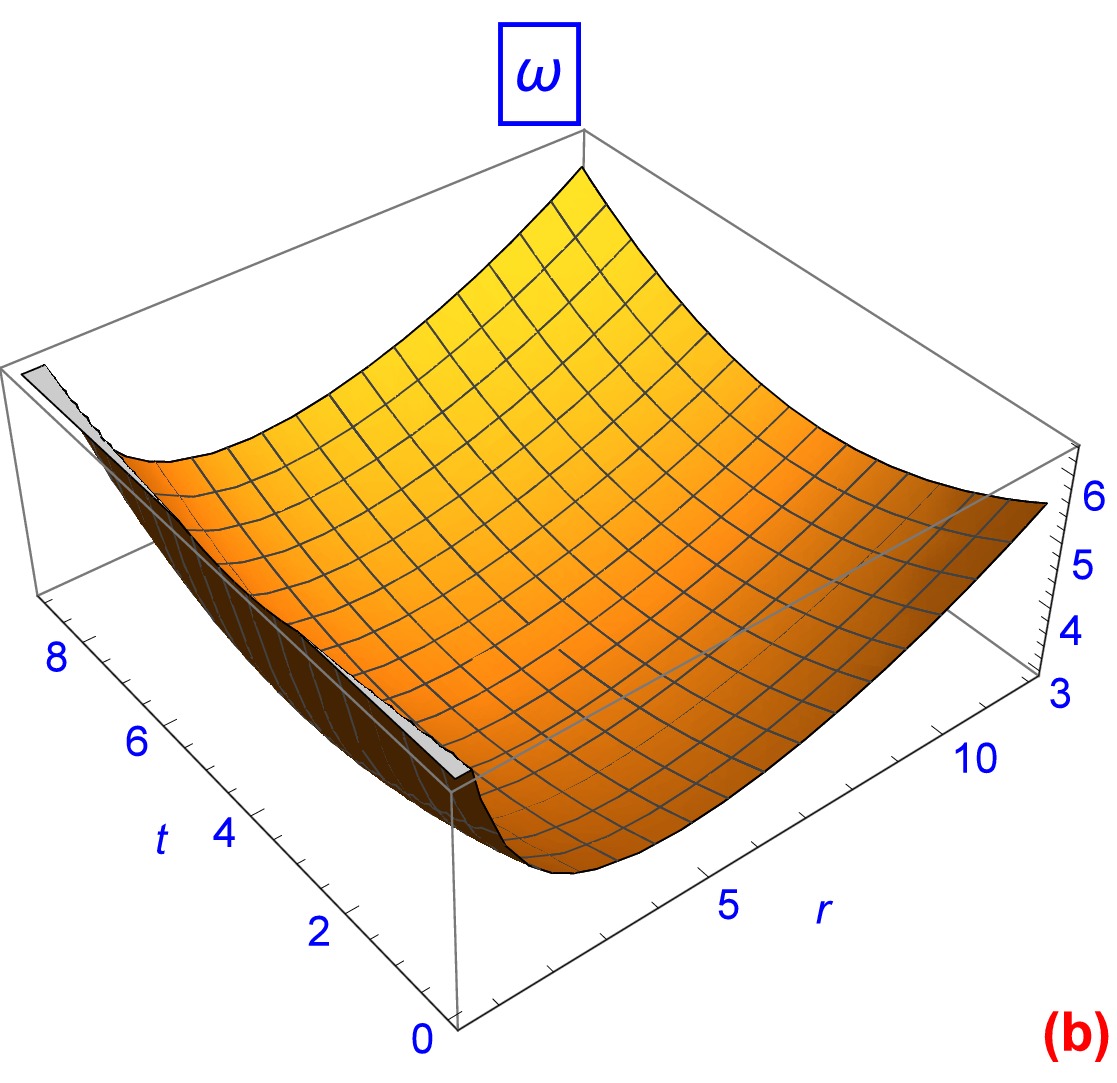}}
\centerline{
\includegraphics[width=5cm]{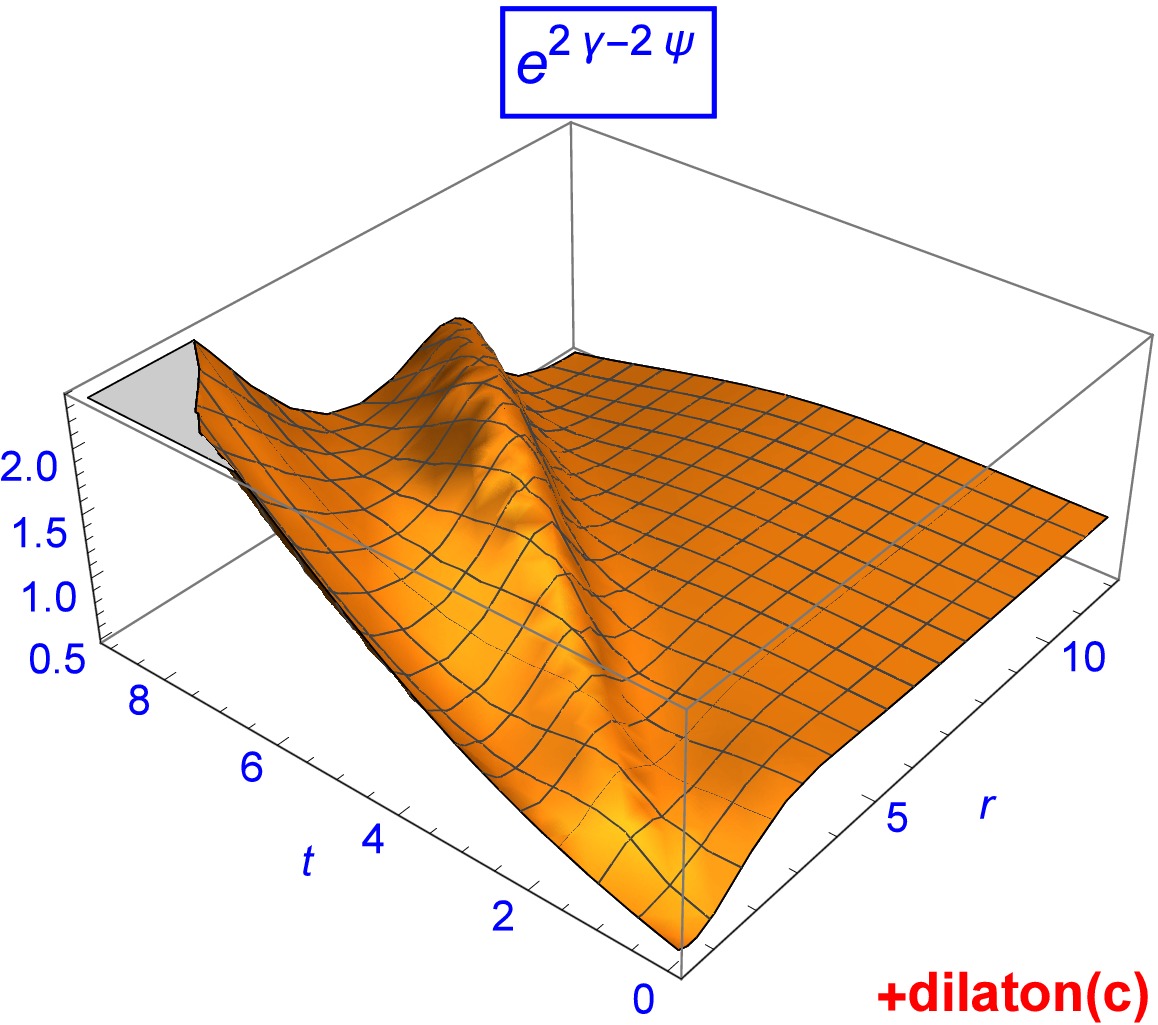}
\includegraphics[width=5cm]{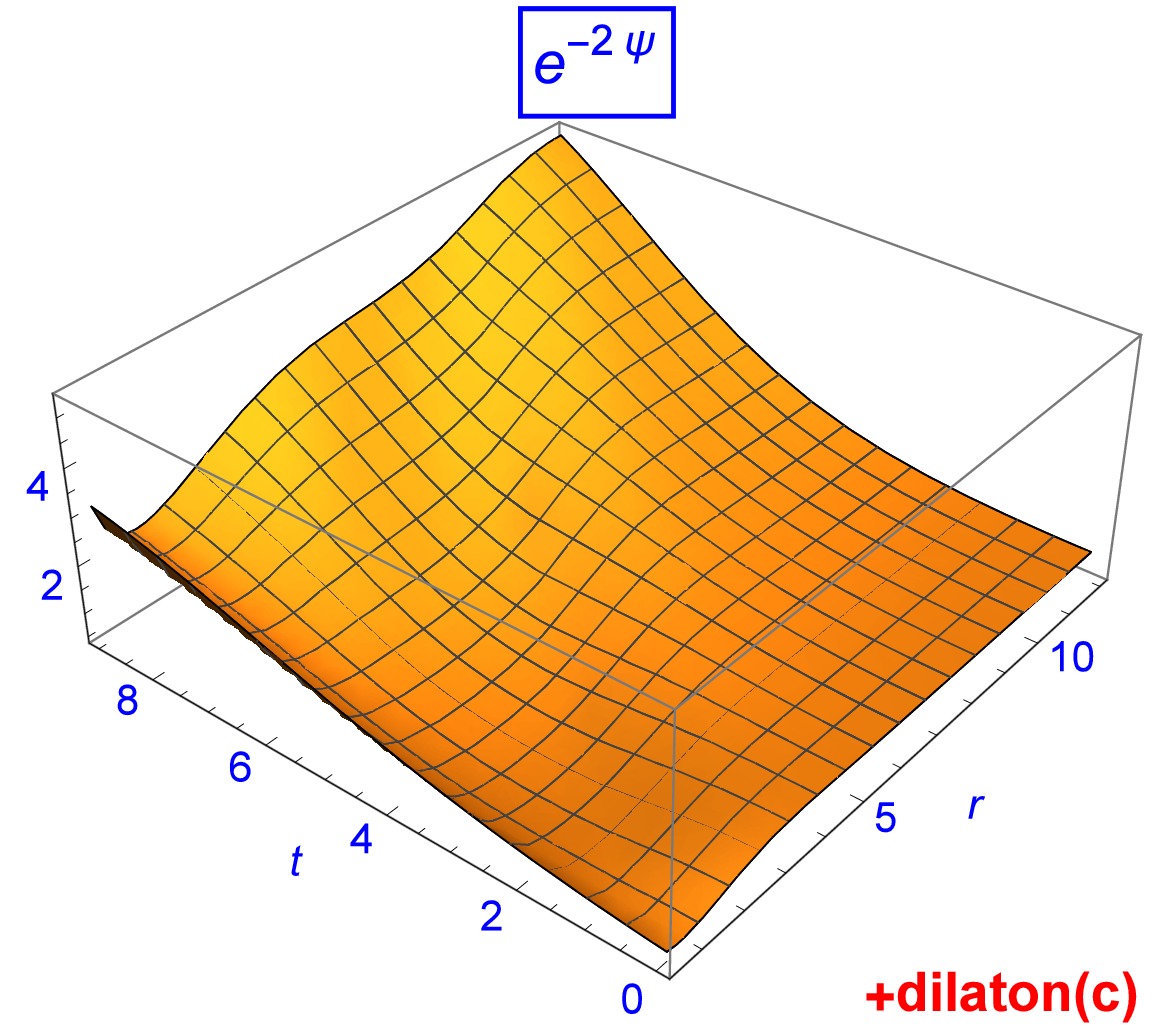}
\includegraphics[width=5cm]{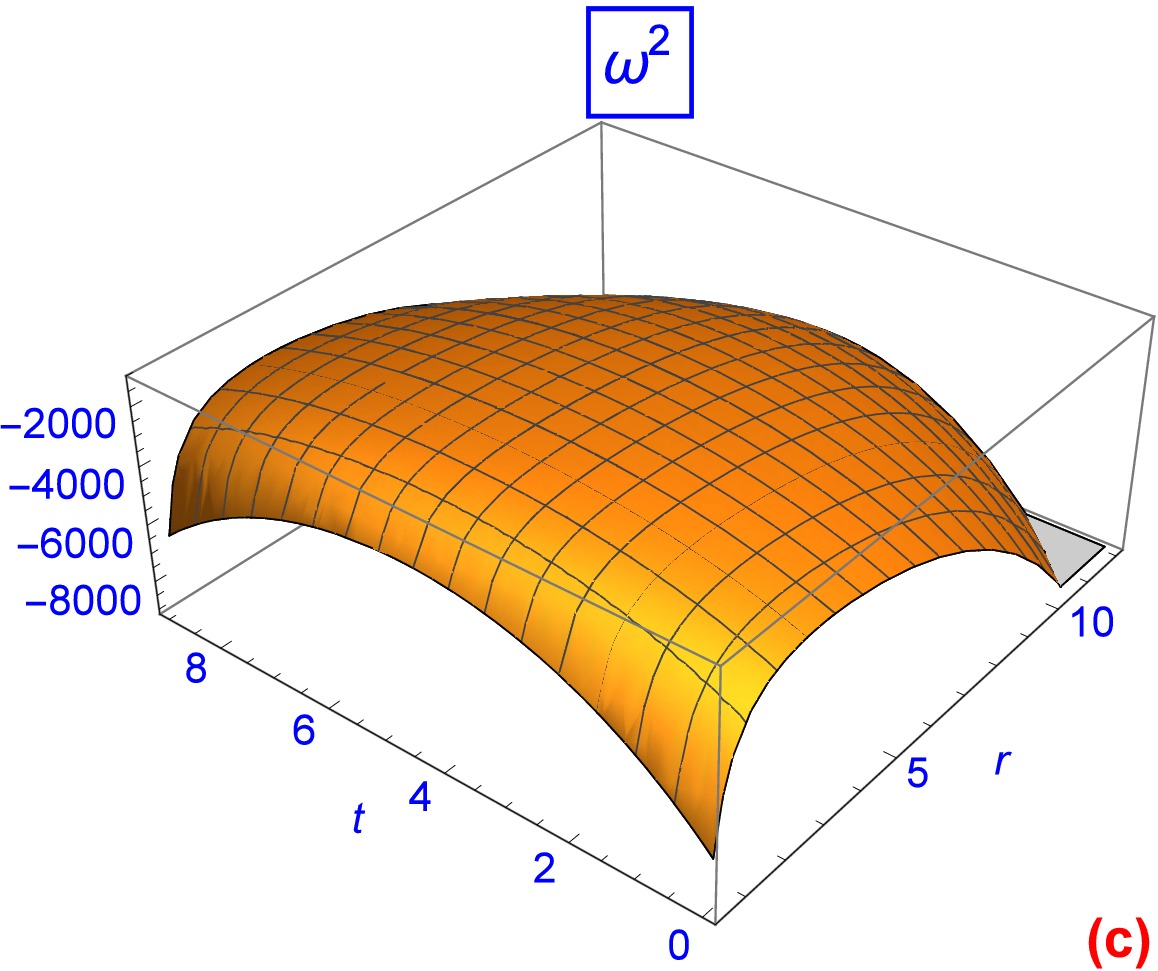}}
\centerline{
\includegraphics[width=5cm]{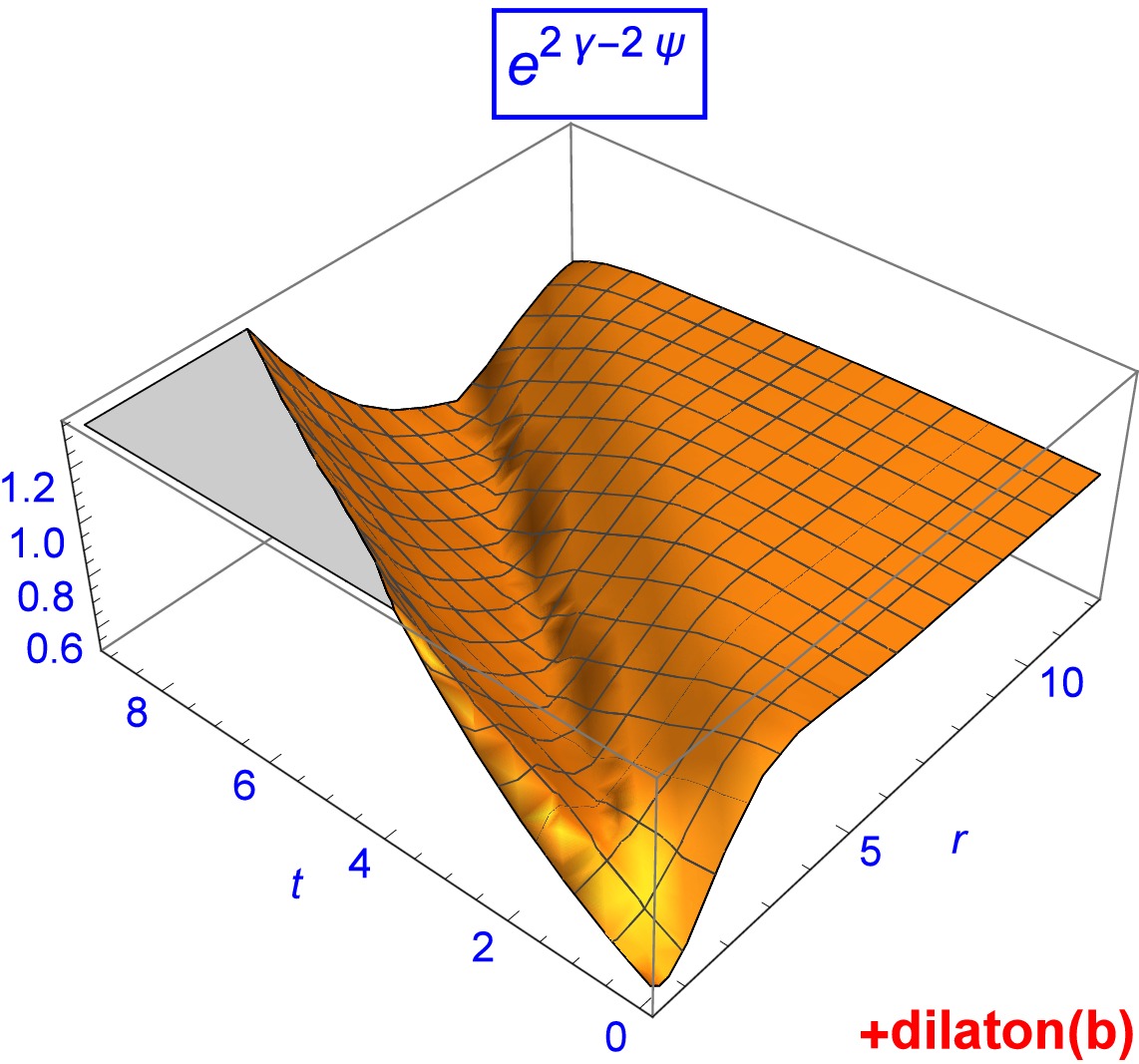}
\includegraphics[width=5cm]{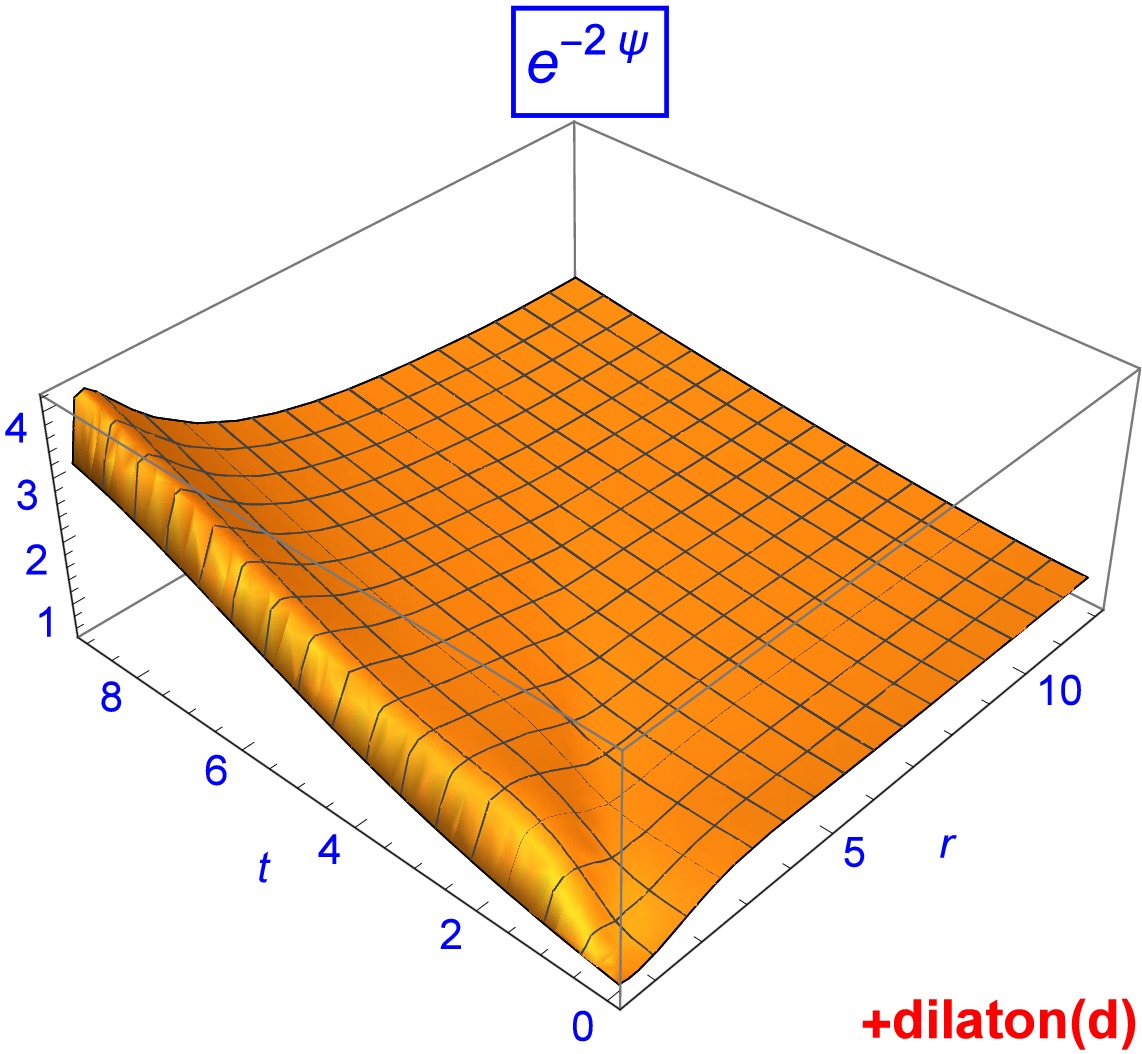}
\includegraphics[width=5cm]{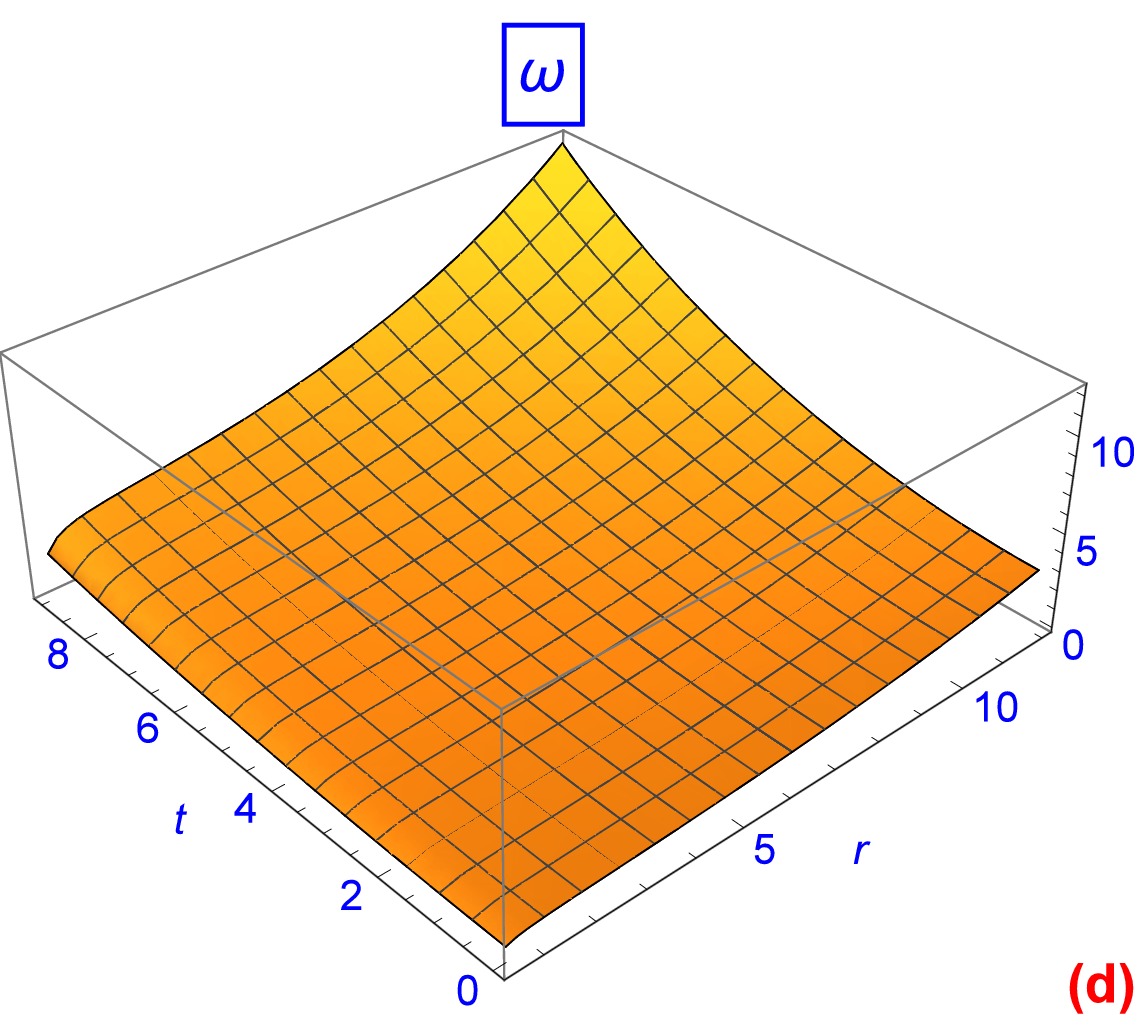}}
\caption{Typical solutions of the un-physical metric $\tilde g_{\mu\nu}$ (Eq.(\ref{eqn20})-Eq.(\ref{eqn21})) for some values of $\tau$ and $d_i$ for $\omega$( Eq.(\ref{eqn10})) compared with the classical Weyl solution of Eq.(\ref{eqn23}) (top). For $\psi$ we took as initial condition a Weber-Wheeler pulse wave.  We observe that in some cases the metric becomes asymptotically flat. For situation (c) is $\omega$ complex.}
\end{figure}
Written out in components, the equations for $\tilde\gamma$ and $\tilde\psi$ become
\begin{eqnarray}
\partial_{tt}\tilde\gamma =\partial_{rr}\tilde\gamma+(\partial_r\tilde\psi)^2-(\partial_t\tilde\psi)^2+\frac{3}{\omega}\Bigl(\partial_r\omega\partial_r \tilde\psi-\partial_t\omega\partial_t\tilde\psi-\frac{\partial_r\omega}{2r} \Bigr),\label{eqn20}
\end{eqnarray}
\begin{eqnarray}
\partial_{tt}\tilde\psi=\partial_{rr}\tilde\psi+\frac{\partial_r\tilde\psi}{r}+\frac{3}{\omega}\Bigl(\partial_r\omega\partial_r \tilde\psi-\partial_t\omega\partial_t\tilde\psi-\frac{\partial_r\omega}{2r} \Bigr).\label{eqn21}
\end{eqnarray}
The equation for $\omega$ cannot be isolated from the effective 4D Einstein equations, as was already concluded in\cite{slag1}. So we will use the dilaton equation Eq.(\ref{eqn9}).
From the equations Eq.(\ref{eqn9}), Eq.(\ref{eqn13}) and Eq.(\ref{eqn19}) we obtain the constraint equation
\begin{eqnarray}
(\partial_t\omega)^2-(\partial_r\omega)^2-\frac{\omega}{r}\partial_r\omega =0.\label{eqn22}
\end{eqnarray}
We must note that in the non-vacuum model\cite{slag1}, the dilaton equation Eq.(\ref{eqn9}) was obtained by the embedding of the 4D spacetime into a 5D warped spacetime. However, the constraint equations  Eq.(\ref{eqn22}) can become different.

We should like to compare the solution for $\tilde g_{\mu\nu}$ with the "classical" vacuum axially symmetric Weyl solution of the system (the subscript w stands for Weyl)
\begin{eqnarray}
\partial_t\gamma_w=2r\partial_r\psi_w\partial_t\psi_w,\quad \partial_r\gamma_w=r\Bigl((\partial_r\psi_w)^2+(\partial_t\psi_w)^2\Bigr),\quad \partial_{tt}\psi_w=\partial_{rr}\psi_w+\frac{\partial_r\psi_w}{r}.\label{eqn23}
\end{eqnarray}
One can solve the Laplace equation for $\psi_w$ and then integrate the first-order equations for $\gamma_w$.  An integrability condition  follows from $R=0$.
However, many solutions can quickly be obtained from the stationary axially symmetric counterpart model by the substitution $t\rightarrow iz, z\rightarrow it$\cite{step}.
An example is the Einstein-Rosen spacetime.
The advantage of working in this axially symmetric coordinate system is the possibility to generate new solutions (for example the "electro-vac" solution) from  vacuum solutions. If there is rotation, one can use the method of Ernst\cite{islam}, to generate the well known Kerr solution. Moreover, these axially symmetric models admit radiation effects, even in (conformally) flat spacetimes.
Further, non-conform-stationary (vacuum) solutions are only possible in axially symmetric models\cite{perj}.
A spherical mass surrounded by empty space is truly isolated, but a cylindrical mass distribution will cause energy flow to and from infinity.  If an initially static solution emits a pulse of radiation, then there will be a change in  the values of the  parameters describing the solution (Birkhoff's theorem).

A well studied solution of Eq.(\ref{eqn23}) is the (complex) Weyl solution ($z\rightarrow it$)
\begin{eqnarray}
\Psi_w =C\ln  \left( {\frac {\sqrt {{r}^{2}+ \left( it-m \right) ^{2}}+\sqrt {{r}^{2}+ \left( it+m \right) ^{2}}-2\,m}{\sqrt {{r}^{2}+
\left( it-m \right) ^{2}}+\sqrt {{r}^{2}+ \left( it+m \right) ^{2}}+2\,m}} \right)\label{eqn24}
\end{eqnarray}
This solution leads for $C=1$ to the Schwarzschild metric by the transformation \\ $r\rightarrow\sqrt{r^2-2mr}\sin\theta, z=(r-m)\cos\theta$.
In section 6 we will use a slightly different solution of Eq.(\ref{eqn23}) for our conformal invariant model.

In figure 2 we plotted a typical solution for $e^{2\tilde\gamma}$ and $e^{2\tilde\psi}$. We also plotted, for comparison, the "classical" Weyl solution. For $\omega$ we took a typical value from Eq.(\ref{eqn10}).
It is obvious that the dilaton plays a crucial role in the evolution of the metric.
It is a foretaste of the cosmological significancy of $\omega$.

We can apply the Cauchy-Kowalewski theorem for conformal invariancy: we can choose Eq.(\ref{eqn22}) as initial condition on the boundary of $\tilde g_{\mu\nu}$.
If we substitute Eq.(\ref{eqn22}) into Eq.(\ref{eqn9})  we obtain the Laplace equation and next from Eq.(\ref{eqn13}) we see that $\tilde R=0$, i.e., the flat spacetime.
One could also say that our dilaton equation Eq.(\ref{eqn9}) can conformally transformed ($\omega\rightarrow\frac{1}{\Omega}\omega$)   into the Laplace equation in order that Eq.(\ref{eqn22}) holds.
This yields a gauge condition for $\Omega$.
This result is also formulated as follows: the vacuum  $R=0$ breaks local conformal invariance, unless we impose  Eq.(\ref{eqn9}) together with Eq.(\ref{eqn22}).
We will return to this issue in the next section.
\section{Generation of conformally Ricci-flat $\tilde g_{\mu\nu}$}
\label{sec:5}
To get an indication how to proceed with the un-physical metric $\tilde g_{\mu\nu}$ in order to ends up with a Ricci-flat spacetime by conformal transformations, we
consider, as an  illustrative example, the Minkowski spacetime written in radiative coordinates [see textbook of Wald\cite{wald})
\begin{equation}
ds^2=-dudv+\frac{1}{4}(v-u)^2(d\theta^2+sin^2\theta d\varphi^2),\label{eqn25}
\end{equation}
with $v=t+r, u=t-r$. One needs information about the behavior of fields when $v\rightarrow\infty$. Introducing $V=\frac{1}{v}$ we obtain
\begin{equation}
ds^2=\frac{1}{V^2}\Bigl(dudV+\frac{1}{4}(1-uV)^2(d\theta^2+sin^2\theta d\varphi^2\Bigr),\label{eqn26}
\end{equation}
and "infinity"  corresponds to $V=0$. But the metric is singular at $V=0$. We have obtained a bad coordinate system. Suppose we define an un-physical metric $\tilde g_{\mu\nu}=V^2\eta_{\mu\nu}$. We then obtain a smooth metric extended to $V=0$ and can handle tensor analysis at infinity. One can even do better by introducing the conformal factor $\tilde g_{\mu\nu}=\frac{4}{(1+v^2)(1+u^2)}\eta_{\mu\nu}$. If one chooses the coordinates $T=tan^{-1}v+tan^{-1}u$ and $R=tan^{-1}v -tan^{-1}u$, one obtains the static $ ( S^3 \otimes\Re) $ Einstein universe
\begin{equation}
d\tilde s^2=-dT^2+dR^2+sin^2 R(d\theta^2+sin^2\theta d\varphi^2).\label{eqn27}
\end{equation}
So there exists a conformal map of $(\Re^4,\eta_{\mu\nu})$ into an open region  $( S^3\otimes \Re,\tilde g_{\mu\nu})$ restricted by $-\pi < (T \pm R)< \pi , R\geq 0$. The conformal infinity of Minkowski is then the boundary in the Einstein static universe.

Let us now consider the special Ricci-flat solution of the Weyl class of the equations  Eq.(\ref{eqn23})\cite{har}
\begin{eqnarray}
\psi_w^{(0)}=c_1 \ln \Bigl(t+\sqrt{t^2-r^2}\Bigr),\qquad \gamma_w^{(0)}=2c_1^2\ln \frac{1}{2}\Bigl(\frac{t}{\sqrt{t^2-r^2}}+1 \Bigr)+c_2,\label{eqn28}
\end{eqnarray}
with $c_i$ constants of integration. This Ricci-flat Weyl solution has some interesting properties, i.e., the C-energy is non-vanishing and $c_2\neq 0$ introduces a conical singularity. For some values of $c_1$ are the solutions self-similar. For $c_1=\frac{1}{2}$ it is flat. Let us denote this metric as $g_{\mu\nu}^{(0)}$. In order to maintain Ricci-flat spacetimes  after the conformal map $g_{\mu\nu}^{(0)}\rightarrow \Omega^2 g_{\mu\nu}^{(0)}$, $\Omega$ must satisfy again Laplace equation ( see Eq.(\ref{eqn13}) with $\omega$ replaced by $\Omega$)
\begin{equation}
\partial_{tt}\Omega -\partial_{rr}\Omega-\frac{\partial_r\Omega}{r}=0.\label{eqn29}
\end{equation}
The constraint equation from the Einstein equations is again Eq.(\ref{eqn22}), now  for $\Omega$
\begin{equation}
(\partial_t\Omega)^2-(\partial_r\Omega)^2-\frac{\Omega}{r}\partial_r\Omega =0\label{eqn30}
\end{equation}
and doesn't contain the constants $c_i$. The constant $c_1$ enters an initial condition for $\partial_r\Omega$. A special solution is $\Omega=\frac{1}{\sqrt{t^2-r^2}}$.
So one can construct self-similar Einstein-Rosen spacetimes.
There exist many generating methods to obtain, for example, Einstein-Rosen soliton wave solutions, superimposed on a Levi-Civita seed\cite{vand}.
This generation procedure complicates of course considerable for our $\tilde g_{\mu\nu}$ of Eq.(\ref{eqn8})
In the non-vacuum case  becomes the procedure even worse. Conformal invariance will then be broken.
We will study this problem in the next section.

\section{Matter comes into play}
\label{sec:8}
The reason for writing the FRLW spacetime Eq.(\ref{eqn1}) in polar coordinates will become clear when we  include the U(1) scalar gauge fields  into the model, i.e., Eq.(\ref{eqn7}). In a FLRW spacetime, any spacelike geodesic with $t=$ constant delineates an axis of rotational symmetry, so it is always possible to rotate the coordinate system in such a way that this axis becomes the polar axis.
In a former study\cite{slag1} it was found that in order to incorporate the Nielsen-Olesen vortex solution of the U(1) scalar gauge field (axially symmetric!), one needs the FLRW in polar coordinates. Radiative effects can then be studied and the behavior of the "string-like" matter field. One must consider two regions, i.e., the moment in time when the Hubble radius is much larger than the string-core and the moment in time when  the radius of the vortex was comparable with the Hubble radius.
A nasty problem is that the late-time approximate spacetime is of the form\cite{greg}
\begin{equation}
ds^2=a(t)\Bigl[-dt^2+dr^2+K(r)^2dz^2+(1-4G\mu)^2S(r)^2d\varphi^2\Bigr]\label{eqn31},
\end{equation}
or transformed
\begin{equation}
ds^2=a(t)^2\Bigl[-d\tau^2+\frac{dR^2}{1-kR^2}+R^2d\theta^2+(1-4G\mu)^2R^2\sin^2\theta d\varphi^2\label{eqn32}.
\end{equation}
Some tricky matching conditions are necessary at the boundary with the radiating Einstein-Rosen spacetime $\tilde g_{\mu\nu}$\cite{and}. For $k=-1$ (open universe) these conditions can be found.
The spacetime of  Eq.(\ref{eqn31}) has a residual angle deficit proportional to the mass density of the string.
However, in the warped 5D counterpart model\cite{slag4}, this remnant disappears by the effect of the warp factor. Further, the magnitude of cylindrical gravitational waves is proportional to the ratio of the string core radius $r_s$ and the Hubble radius $R_H$. This ratio $r_s/R_H$ is  negligible at late times but not at the moment of formation:  the vortex builds up a huge mass by the presence of the bulk spacetime. It was conjectured that another effect could  emerge just after the symmetry breaking phase of the scalar gauge field: preferred azimuthal angle of the spinning axes of quasars in large quasar groups{\cite{slag2,slag3}.
These considerations can now be related to the conformal invariance approach.

The resemblance of equations for the dilaton $\omega$ and the scalar field $\Phi$ of  Eq.(\ref{eqn13}) and  Eq.(\ref{eqn14}) (in 4D)  is evident.
However, matter fields must be placed in the Lagrangian for matter fields ${\cal L}_M$ and we must have local energy-momentum conservation $\nabla^\mu T_{\mu\nu}^{(M)}=0$, where
\begin{eqnarray}
{T^{\mu\nu(M)}}=\frac{2}{\sqrt{-g}}\frac{\delta({\cal L}_M\sqrt{-g})}{\delta g_{\mu\nu}}=2\frac{\delta{\cal L}_M}{\delta g_{\mu\nu}}+g^{\mu\nu}{\cal L}^M.\label{eqn33}
\end{eqnarray}
Indeed, our world is not vacuum, so the task is to add a matter Lagrangian ${\cal L}_M$ to the action and investigate if it can be made conformal invariant. Moreover, $T_{\mu\nu}$ must be  traceless\cite{wald}.
When gravity is coupled to matter, it is believed that conformal invariance ( which is an exact local invariance) must be spontaneously broken and will fix all the  parameters of the model.\\
It could be possible that in an eventually conformal invariant gravity theory all physical constants, including Newton's constant, masses and cosmological constant, are in principle computable\cite{thooft5,thooft2,thooft3}.
In the case of our scalar gauge field we have (where we wrote $\Phi\rightarrow\frac{1}{\omega}\tilde\Phi$ and $\tilde\Phi \rightarrow\eta+\frac{\tilde\Phi}{\sqrt{2}}$)
\begin{eqnarray}
{\cal I}=\int d^4x\sqrt{-\tilde g}\Bigl\{\frac{1}{2\kappa_4^2}\Bigl(\omega^2\tilde R+6\partial_\alpha \omega\partial^\alpha \omega-2\Lambda_4 \omega^4\Bigr)-\frac{1}{12}\tilde\Phi\tilde\Phi^*\tilde R \cr-\frac{1}{2}{\cal D}_\alpha\tilde\Phi({\cal D}^\alpha\tilde\Phi)^* -\frac{1}{4}F_{\alpha\beta}F^{\alpha\beta}-V(\tilde\Phi ,\omega)\Bigr\}\label{eqn34}
\end{eqnarray}
where we omitted, for the time being, other interaction terms. The gauge covariant derivative is ${\cal D}_\mu\Phi=\nabla_\mu\Phi+ieA_\mu\Phi$. The term $\frac{1}{12}\tilde R\tilde\Phi\tilde\Phi^*$  makes  the scalar Lagrangian conformal invariant. The resemblance between the two parts in Lagrangian Eq.(\ref{eqn34}) is clear when one redefines $\bar\omega^2\equiv -\frac{6\omega^2}{\kappa_4^2}$  (apart from the potential term $V(\Phi,\omega)$ ). So $\bar\omega$ is complex (unitarity gauge\cite{thooft1}). In this case, this means that now the $\bar\omega^2$-solution is of the form of figure 1-c.
The functional integration over the $\omega$ degree of freedom should be rotated in the complex plane (Wick rotation).
In flat spacetimes this is comparable with the transformation $t\rightarrow it$, which makes the metric positive definite on an analytically extended manifold.
In GR this is problematic, since time has no physical meaning in GR.
Moreover, it can produce complex solutions, for example, the deSitter spacetime. It must be noted that in our axially symmetric spacetime, we defined $t\rightarrow iz$ together with $z\rightarrow it$, so the latter problem doesn't occur.
A more detailed analysis of this issue can be found, for example, in\cite{park}.
Note that the potential $V(\tilde\Phi,\omega)$ is also $\omega$-dependent. We can take for example $V(\tilde\Phi ,\bar\omega)=\frac{1}{8}\beta\eta^2\kappa_4^2\tilde\Phi\tilde\Phi^*\bar\omega^2$ (where the "double-well"-potential mass parameter is now $\beta\eta^2\kappa_4^2\bar\omega^2$).

The re-scaled Lagrangian Eq.(\ref{eqn34}) becomes
\begin{eqnarray}
{\cal I}=\int d^4x\sqrt{-\tilde g}\Bigl\{-\frac{1}{12}\Bigl(\tilde\Phi\tilde\Phi^*+\bar\omega^2\Bigr)\tilde R-\frac{1}{2}\Bigl({\cal D}_\alpha\tilde\Phi({\cal D}^\alpha\tilde\Phi)^*+\partial_\alpha\bar\omega\partial^\alpha\bar\omega\Bigr)\cr
-\frac{1}{4}F_{\alpha\beta}F^{\alpha\beta}-V(\tilde\Phi ,\omega)-\frac{1}{36}\kappa_4^2\Lambda_4\bar\omega^4\Bigr\}\label{eqn35}
\end{eqnarray}
After variation with respect to the field variables, one obtains the equations of motion
\begin{equation}
\tilde G_{\mu\nu}=\frac{1}{(\bar\omega^2 +\tilde\Phi\tilde\Phi^*)}\Bigl(T_{\mu\nu}^{(\bar\omega)}+T_{\mu\nu}^{(\tilde\Phi,c)}+\tilde T_{\mu\nu}^{(A)}+\frac{1}{6}\tilde g_{\mu\nu}\Lambda_{eff}\kappa_4^2\bar\omega^4+\kappa_5^4{\cal S}_{\mu\nu}+\tilde g_{\mu\nu}V(\tilde\Phi,\bar\omega)\Bigr)-{\cal E}_{\mu\nu}\label{eqn36}
\end{equation}
\begin{eqnarray}
\tilde\nabla^\alpha \partial_\alpha\bar\omega -\frac{1}{6}\tilde R\bar\omega -\frac{\partial V}{\partial \bar\omega}-\frac{1}{9}\Lambda_{4} \kappa_4^2\bar\omega^3=0 \label{eqn37}
\end{eqnarray}
\begin{eqnarray}
{\cal D}^\alpha {\cal D}_\alpha\tilde\Phi-\frac{1}{6}\tilde R\tilde\Phi-\frac{\partial V}{\partial\tilde\Phi^*}=0, \qquad \tilde\nabla^\nu F_{\mu\nu}=\frac{i}{2}\epsilon \Bigl(\tilde\Phi ({\cal D}_\mu\tilde\Phi)^* -\tilde\Phi^* {\cal D}_\mu\tilde\Phi\Bigr)\label{eqn38}
\end{eqnarray}
with
\begin{eqnarray}
\tilde T_{\mu\nu}^{(A)}=F_{\mu\alpha}F_\nu^\alpha-\frac{1}{4}\tilde g_{\mu\nu}F_{\alpha\beta}F^{\alpha\beta}\label{eqn39}
\end{eqnarray}
\begin{eqnarray}
\tilde T_{\mu\nu}^{(\tilde\Phi ,c)}=\Bigl(\tilde\nabla_\mu\partial_\nu\tilde\Phi\tilde\Phi^*-\tilde g_{\mu\nu}\tilde\nabla_\alpha\partial^\alpha\tilde\Phi\tilde\Phi^*\Bigr)\cr
-6\Bigl(\frac{1}{2}({\cal D}_\mu\tilde\Phi({\cal D}_\nu\tilde\Phi)^*+({\cal D}_\mu\tilde\Phi)^*{\cal D}_\nu\tilde\Phi)-\frac{1}{2}\tilde g_{\mu\nu}{\cal D}_\alpha\tilde\Phi({\cal D}^\alpha\tilde\Phi)^*\Bigl)\label{eqn40}
\end{eqnarray}
\begin{eqnarray}
T_{\mu\nu}^{(\bar\omega)}=\Bigl(\tilde\nabla_\mu\partial_\nu\bar\omega^2-\tilde g_{\mu\nu}\tilde\nabla_\alpha\partial^\alpha\bar\omega^2\Bigr)-6\Bigl(\partial_\mu\bar \omega\partial_\nu\bar \omega-\frac{1}{2}\tilde g_{\mu\nu}\partial_\alpha\bar \omega\partial^\alpha\bar\omega)\Bigl)\label{eqn41}
\end{eqnarray}
Newton's constant reappears in the quadratic interaction term for the scalar field.
The $\Lambda_{eff}\bar\omega^4$-term is more problematic and we will omit it from now on. For other issues, such as renormalizability and the mechanism of the spontaneous breaking of conformal invariance close to the Planck scale, we refer to the discussion on these subjects by 't Hooft\cite{thooft2,thooft3}. When the curvature radius becomes comparable to the Planck length, the correct gravitational action must contain additional terms such as $R^2, R_{\mu\nu}R^{\mu\nu}$, $R_{\mu\nu\sigma\tau}R^{\mu\nu\sigma\tau}$ or combinations of them. They are the result of the back reaction induced by quantum effects. Vacuum polarization effects will spoil conformal invariance of the "classical" theory (trace anomaly). See for example the textbook of Mukhanov et al.\cite{muk} or Parker et al.\cite{park}.

Here we proceed with the field equations Eq.(\ref{eqn36})-Eq.(\ref{eqn41}) and will try to solve them.
If we calculate the trace of the Einstein equations Eq.(\ref{eqn36}) and using Eq.(\ref{eqn37}) and Eq.(\ref{eqn38}), we have the rest term ( "trace-anomaly")
\begin{equation}
\frac{1}{\bar\omega^2+X^2}\Bigl[16\kappa_4^2\beta\eta^2 X^2\bar\omega^2-\kappa_5^4 n^4\Bigl(\frac{(\partial_r P)^2-(\partial_t P)^2}{r^2\epsilon^2}\Bigr)^2 e^{8\tilde\psi-4\tilde\gamma}\Bigr]\label{eqn42}
\end{equation}
This term brakes the conformal invariance.
The cosmological constant of the brane, $\lambda_4$, appears in the conformal breaking term term of Eq.(\ref{eqn38}) in the case of  $\kappa_5^4\equiv\frac{6\kappa_4^2}{\Lambda_4}$ (RS-balance between  the bulk and brane cosmological constant is broken)
So the quadratic term in the energy momentum tensor, ${\cal S}_{\mu\nu}$, could play an important role in the early universe. It depends also on the multiplicity of the scalar field.
There is still a relation between ${\cal E}_{\mu\nu}$ and ${\cal S}_{\mu\nu}$ from the Bianchi identities, $\nabla^\mu{\cal E}_{\mu\nu}=\kappa_5^4\nabla^\mu{\cal S}_{\mu\nu}$, which shows how (1+3) spacetime variations in the matter-radiation on the brane can source Kaluza-Klein modes.

How should we interpret the dilaton equation for $\omega(t,r)$?
The features of our 5D universe depends, after all, on ${^5 g}_{\mu\nu}$, not  ${^4 \tilde g}_{\mu\nu}$.
It is clear that our dilaton field, by the identification as a warp factor, is a real field that acts on the evolution of our universe\cite{slag1} and can play a role in the explanation of the alignment of the polarization axes of quasars on Mpc scales\cite{slag2,slag3}.
We solved the field equations  Eq.(\ref{eqn36})- Eq.(\ref{eqn41}) for $[\tilde g_{\mu\nu},X,P,\omega]$ and can compare the results with the former solutions  of Eq.(\ref{eqn2}) and Eq.(\ref{eqn7})\cite{slag1}.
A typical solution is given in figure 3 by Mathematica and checked by Maple. The solution depends  on the mass-ratio of the scalar and gauge masses, $\frac{m_A^2}{m_{\Phi}^2}=\frac{\epsilon^2}{\beta}$, the boundary conditions, the multiplicity $n$ and of course the dilaton solution by the constants $d_i$.
\begin{figure}[h]
\centerline{
\includegraphics[width=5cm]{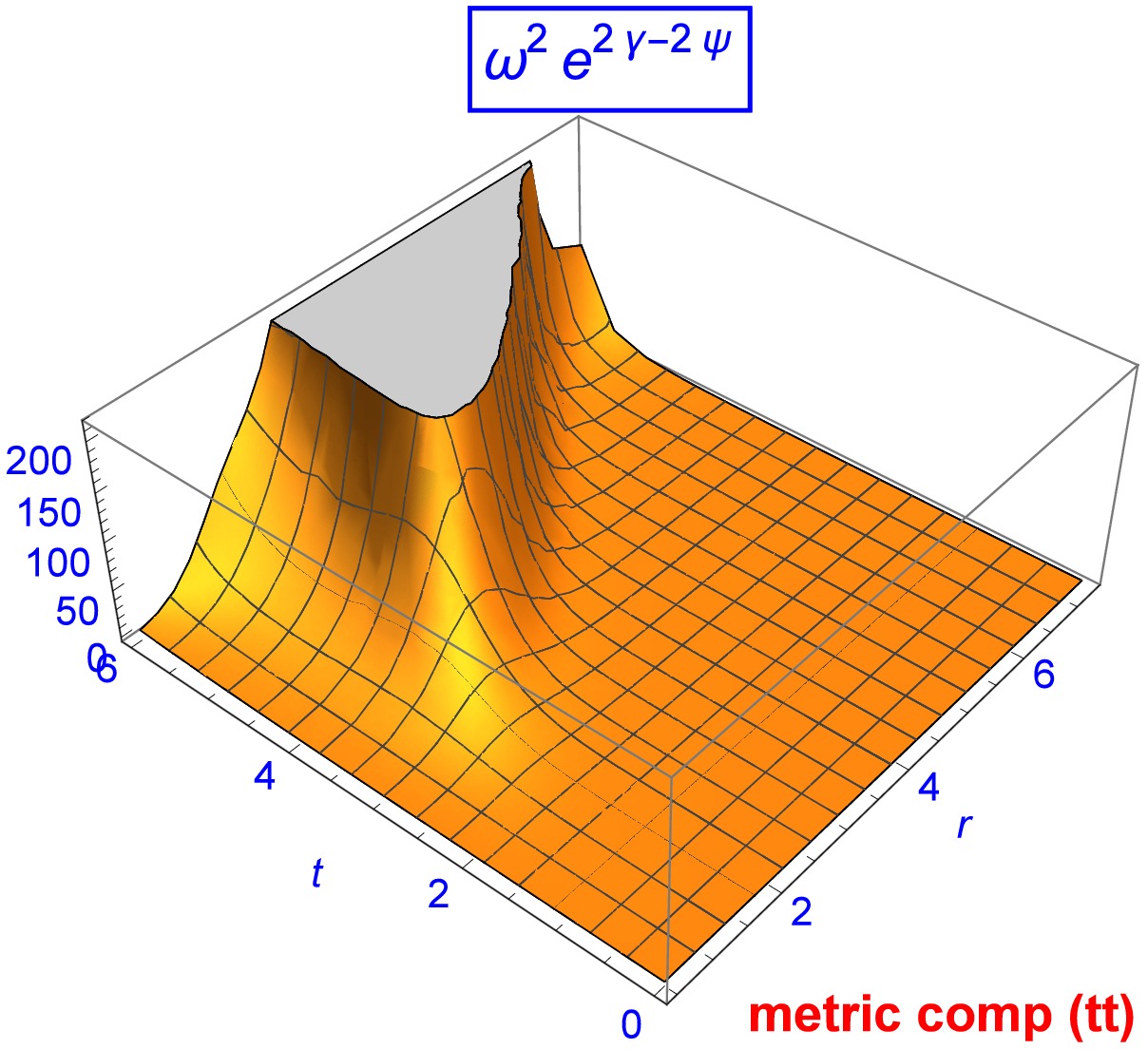}
\includegraphics[width=5cm]{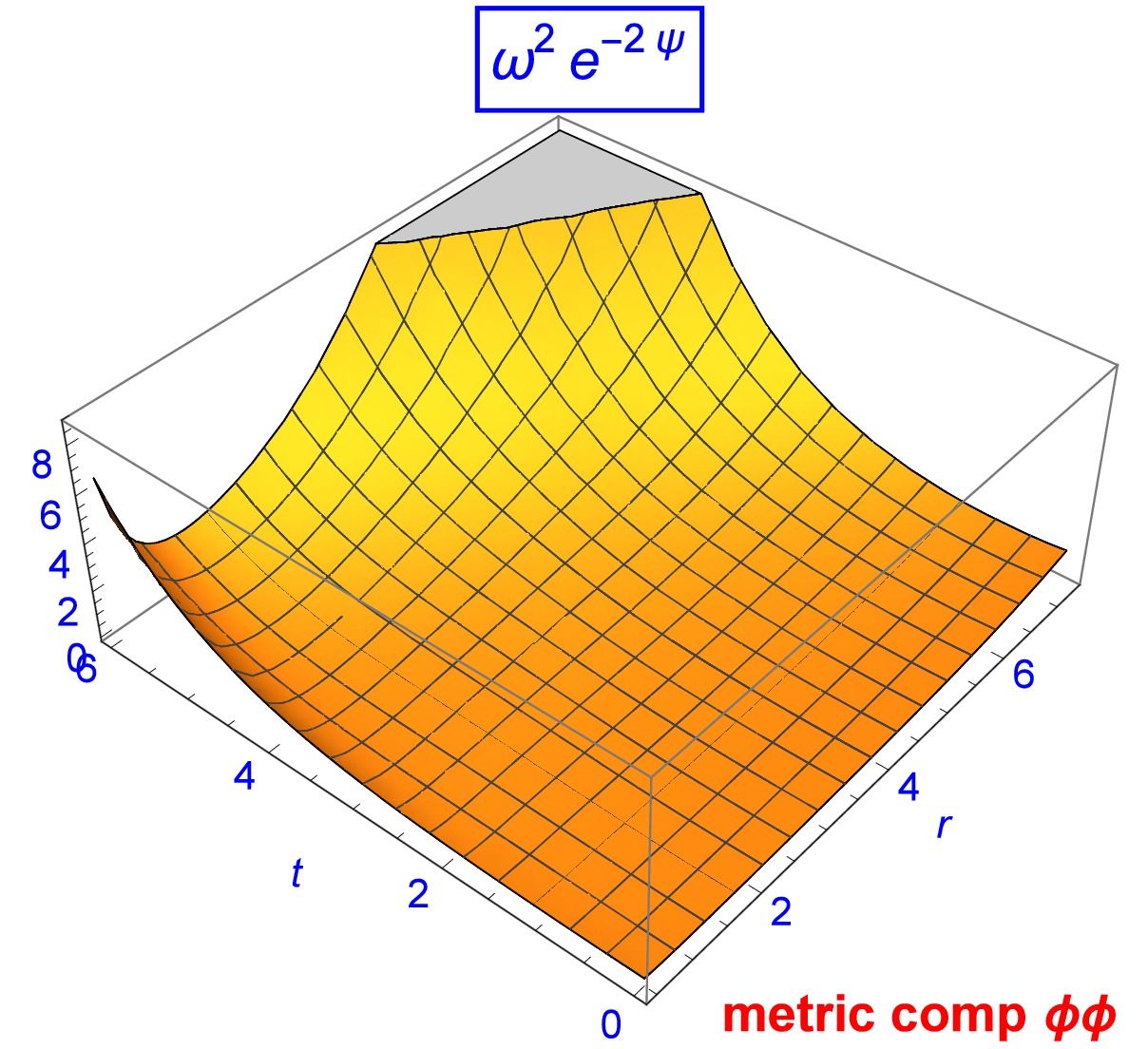}
\includegraphics[width=5cm]{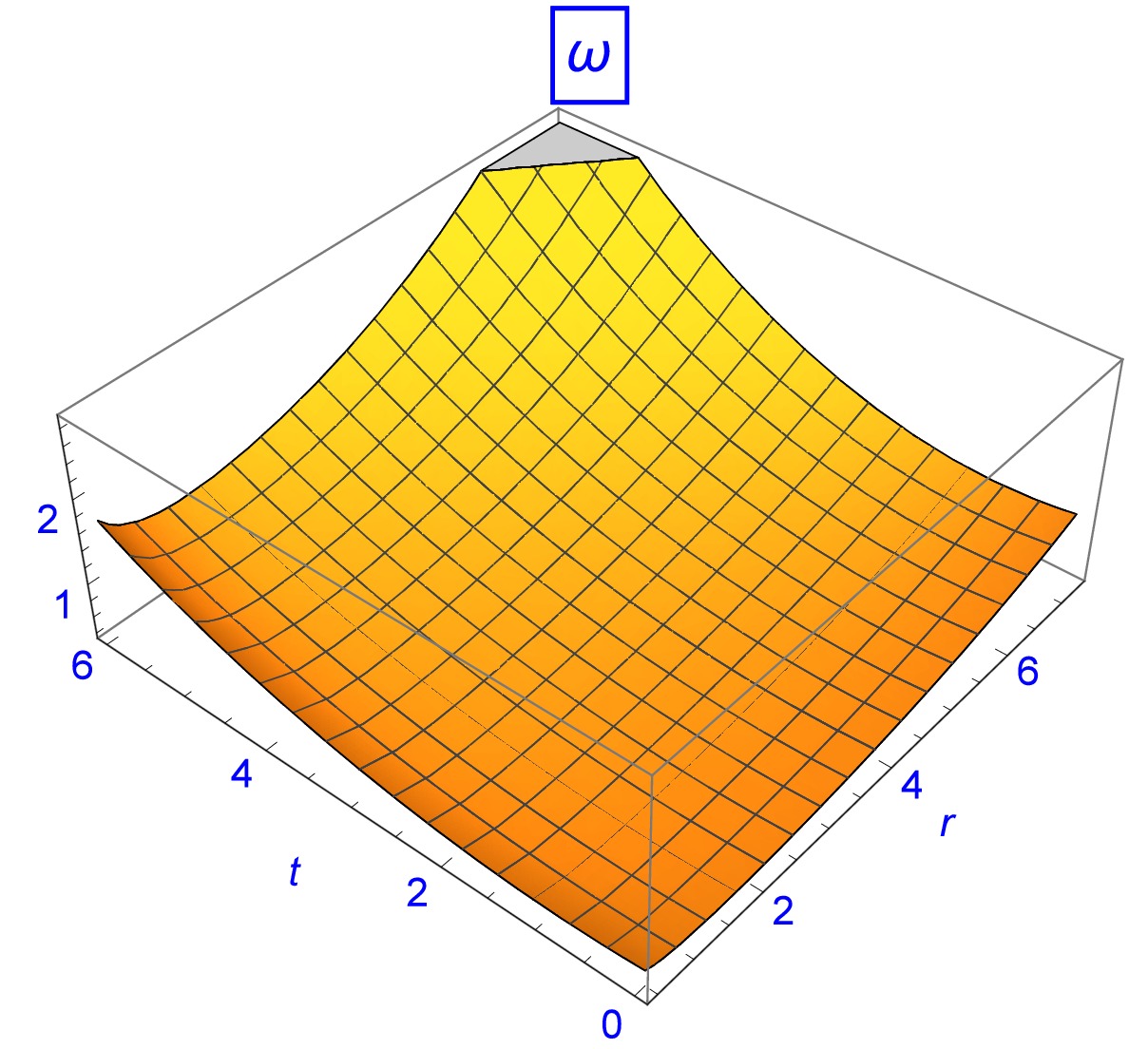}}
\centerline{
\includegraphics[width=5cm]{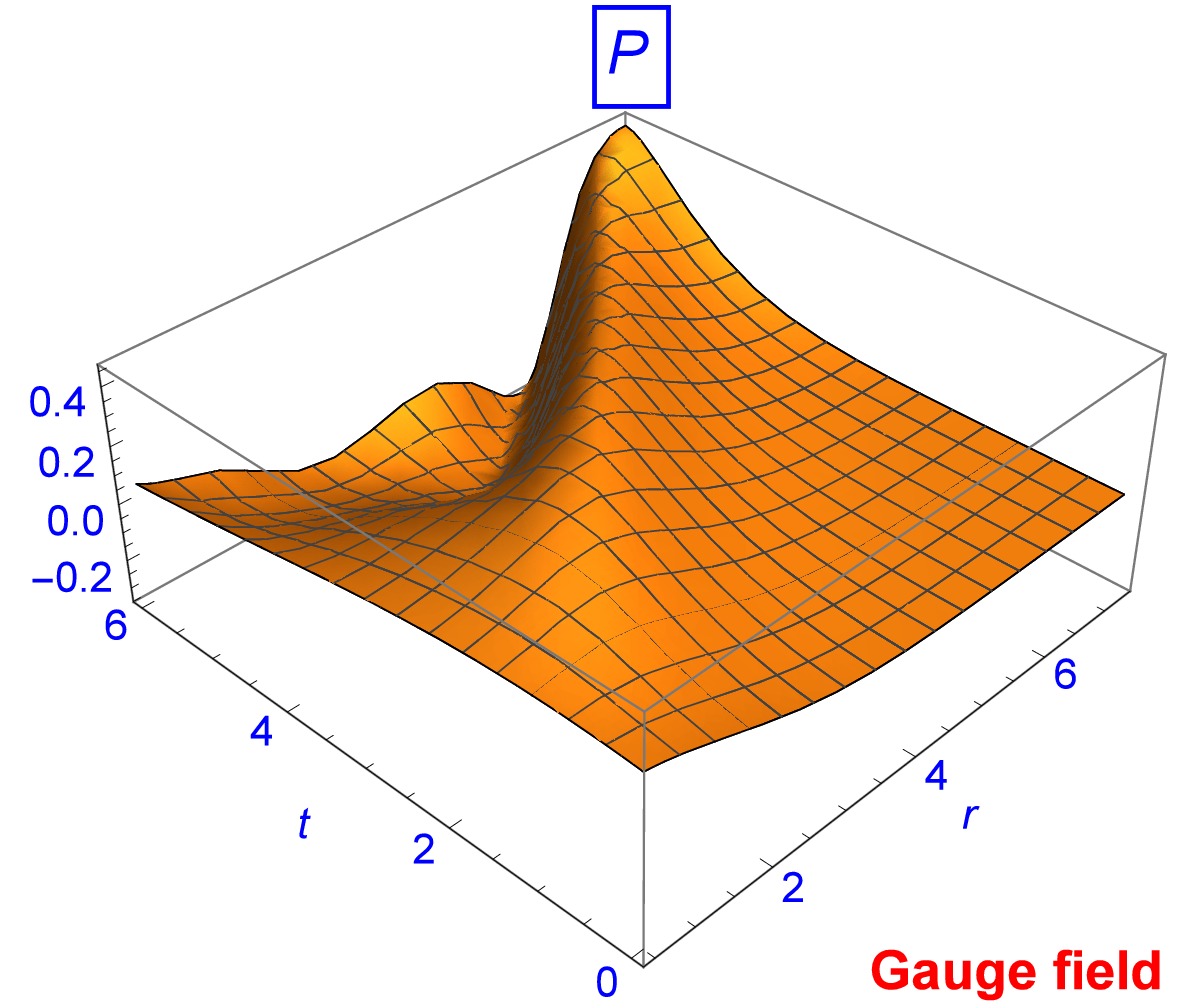}
\includegraphics[width=5cm]{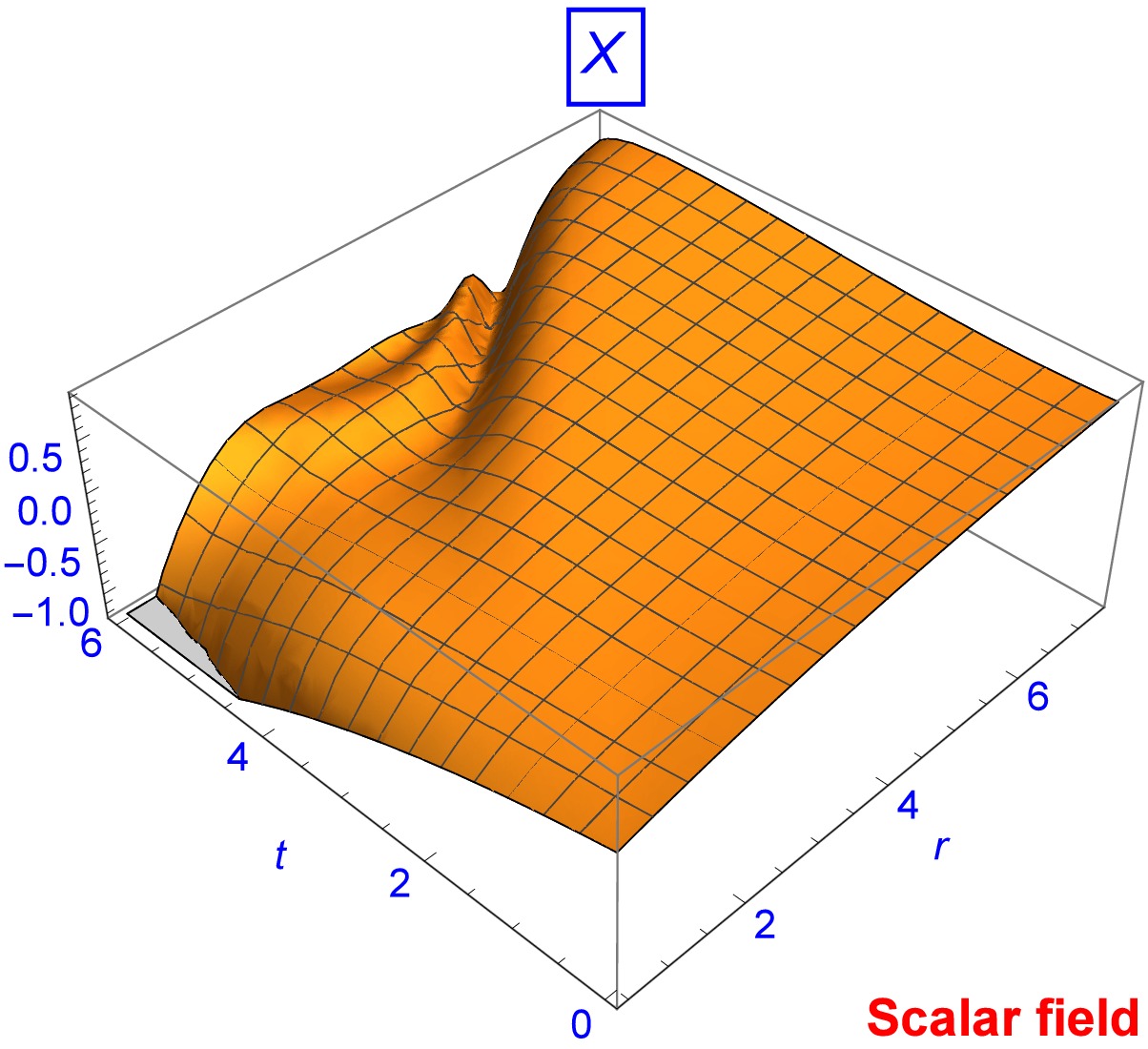}
\includegraphics[width=5cm]{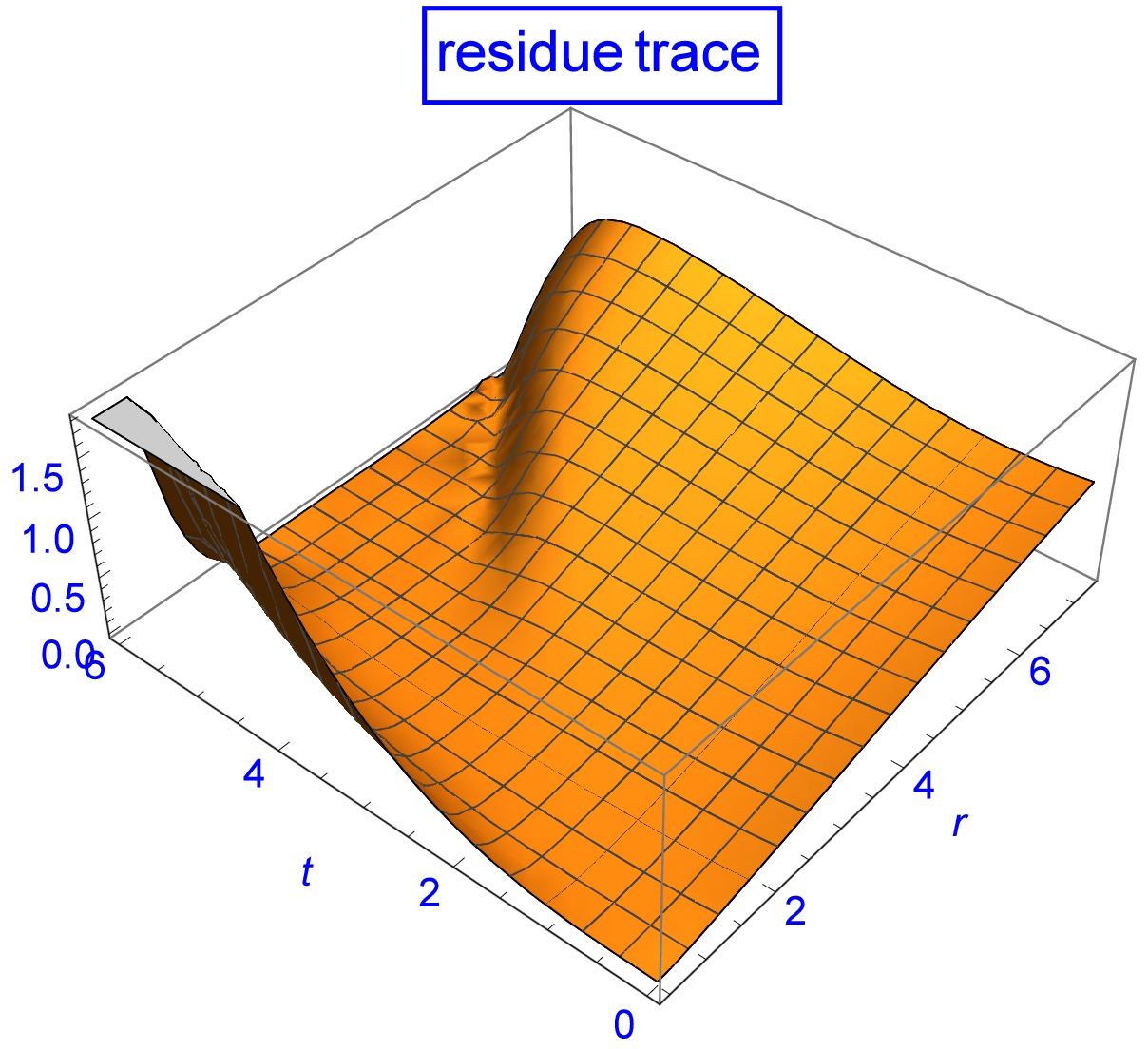}}
\vskip 0.4cm
\caption{Typical solution of the metric of Eq.(\ref{eqn36})- Eq.(\ref{eqn41}). Notice the behavior of the trace of the energy-momentum tensor. We also plotted $\omega$.}
\end{figure}
In figure 4 and 5 we plotted a different solution, with  slightly different values of the parameters.
\begin{figure}[h]
\centerline{
\includegraphics[width=5cm]{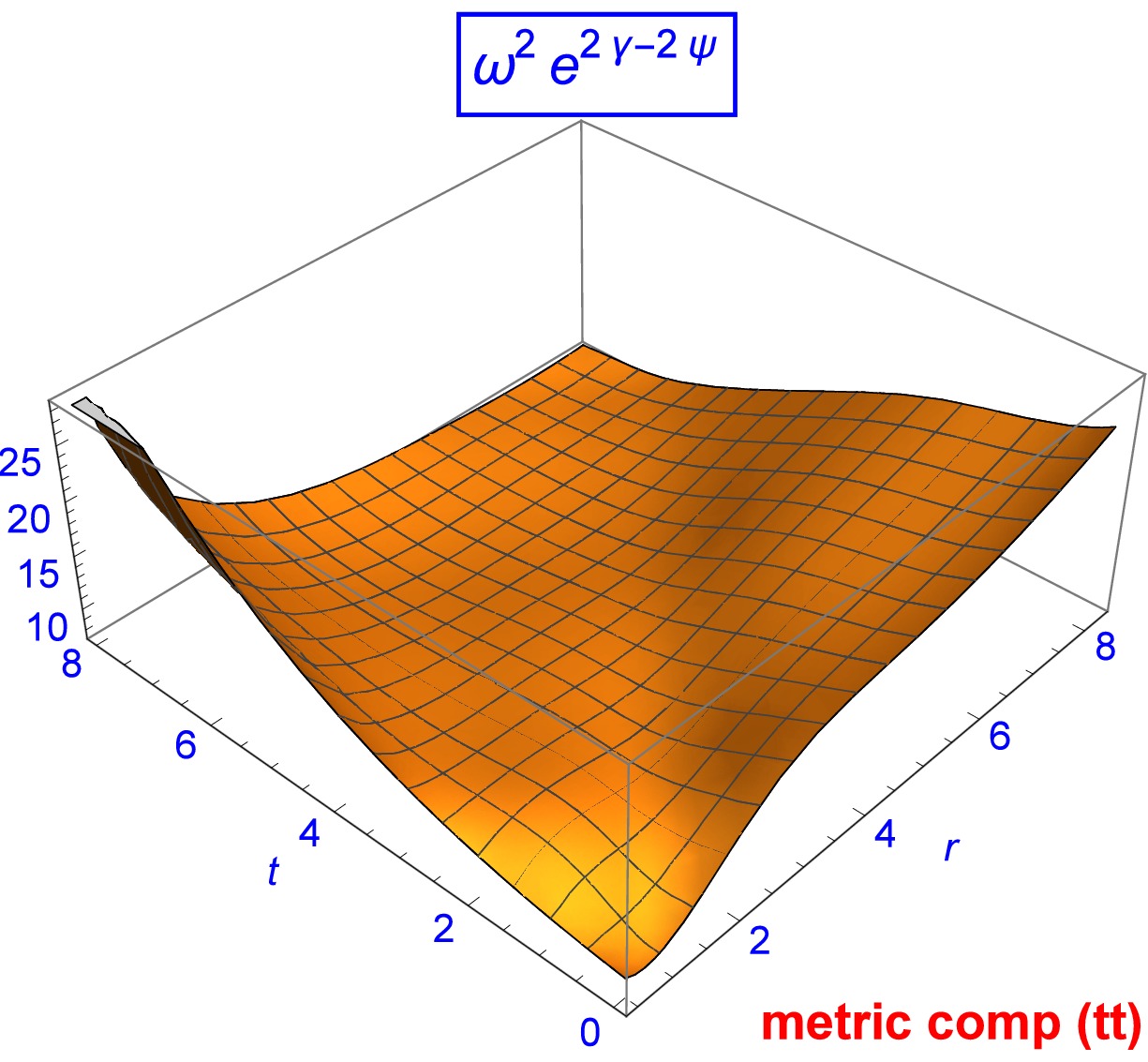}
\includegraphics[width=5cm]{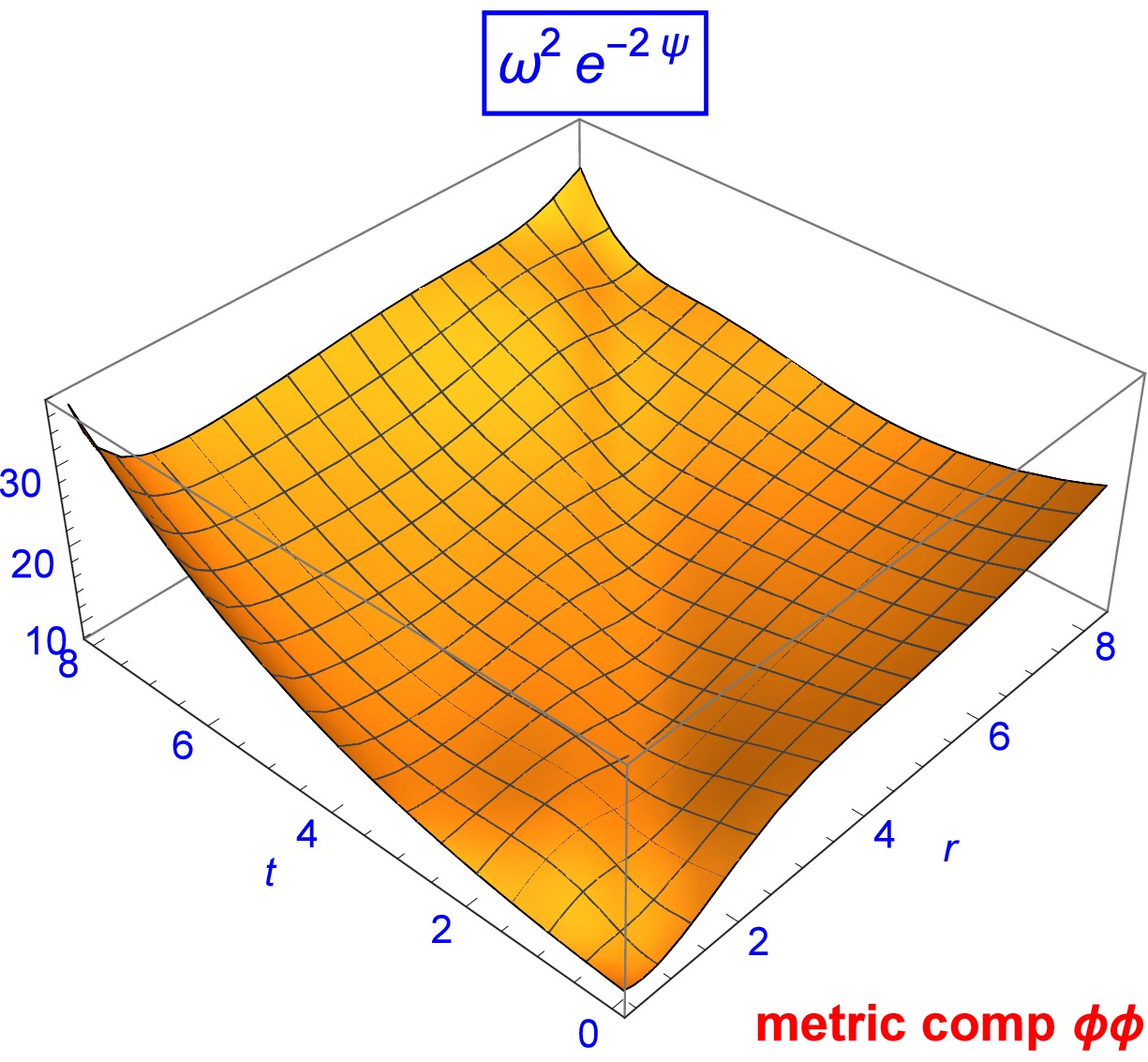}
\includegraphics[width=5cm]{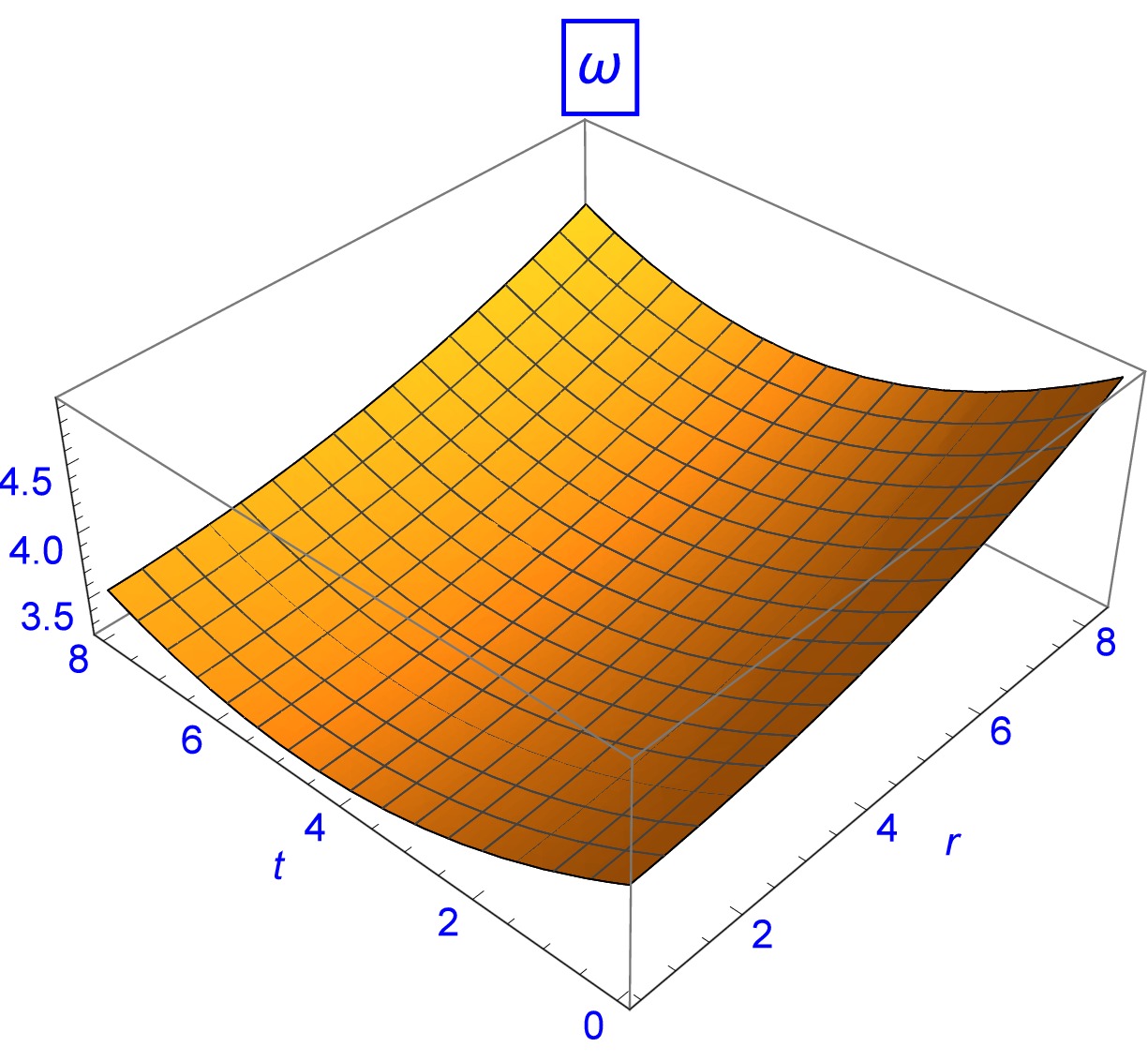}}
\centerline{
\includegraphics[width=5cm]{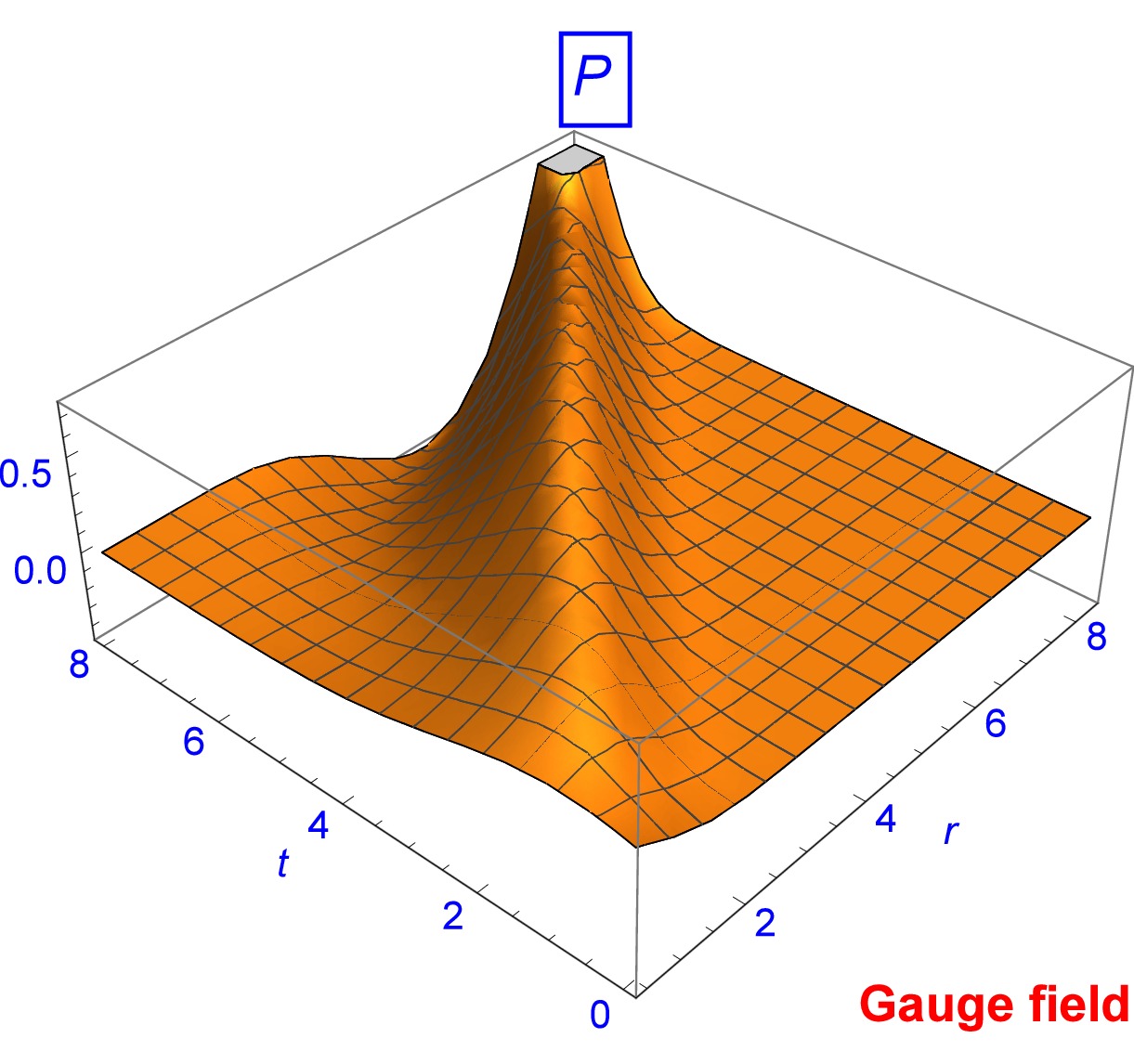}
\includegraphics[width=5cm]{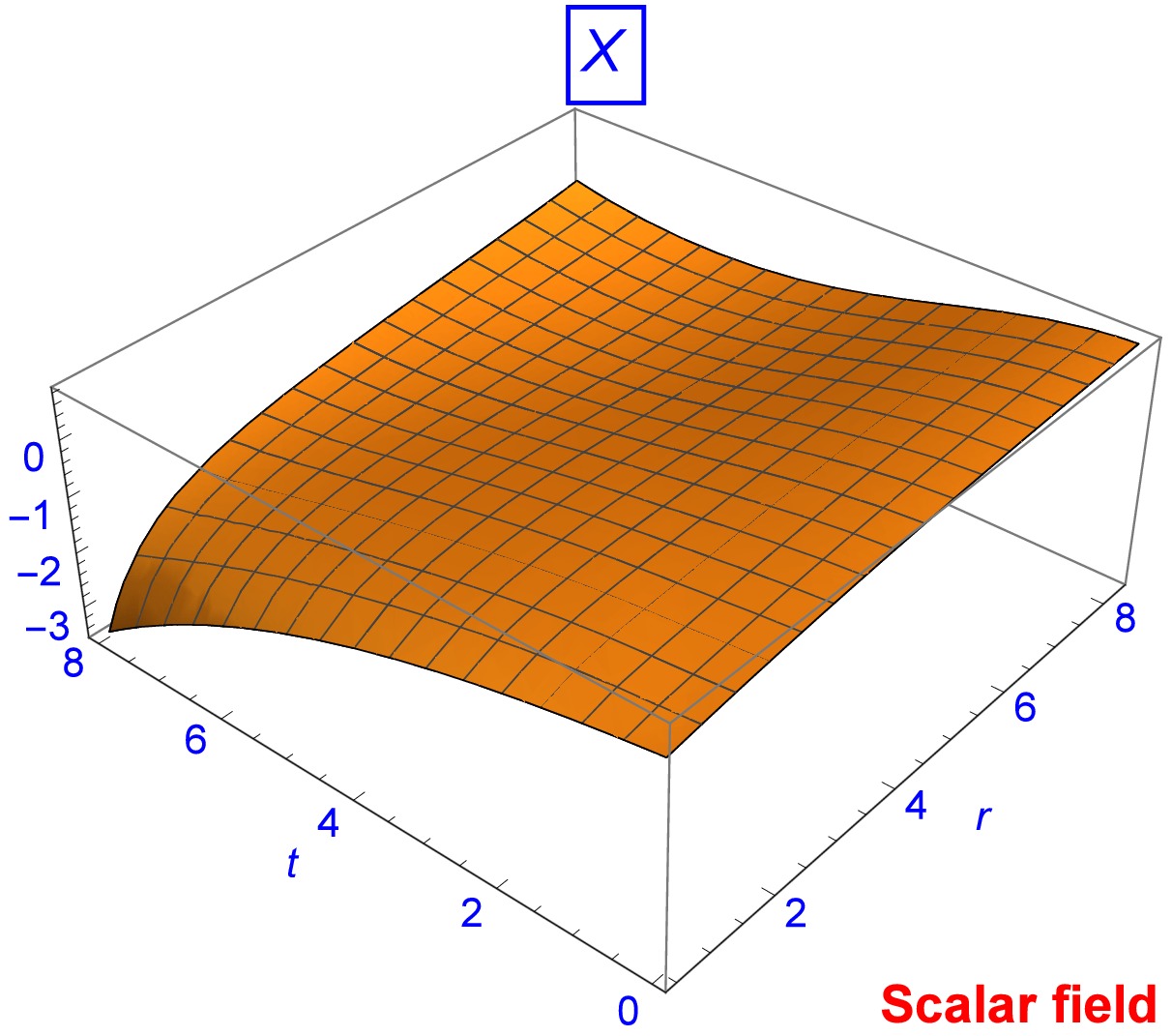}
\includegraphics[width=5cm]{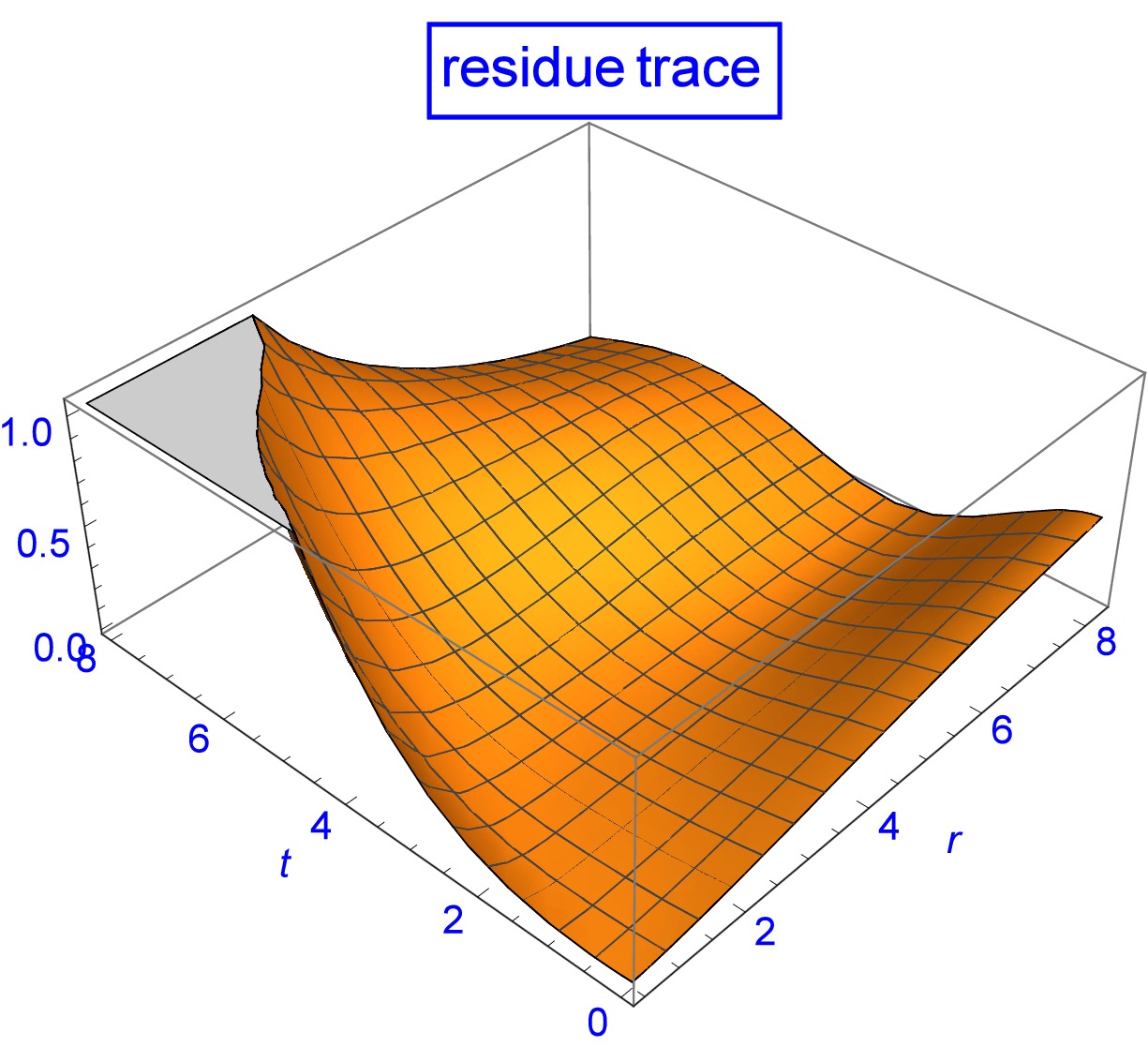}}
\vskip 0.4cm
\caption{As figure 3, for different values of the parameters. The behavior of $g_{tt}$ and $g_{\varphi\varphi}$ differ considerable with respect to the solution of figure 3 }
\end{figure}
\begin{figure}[h]
\centerline{
\includegraphics[width=5cm]{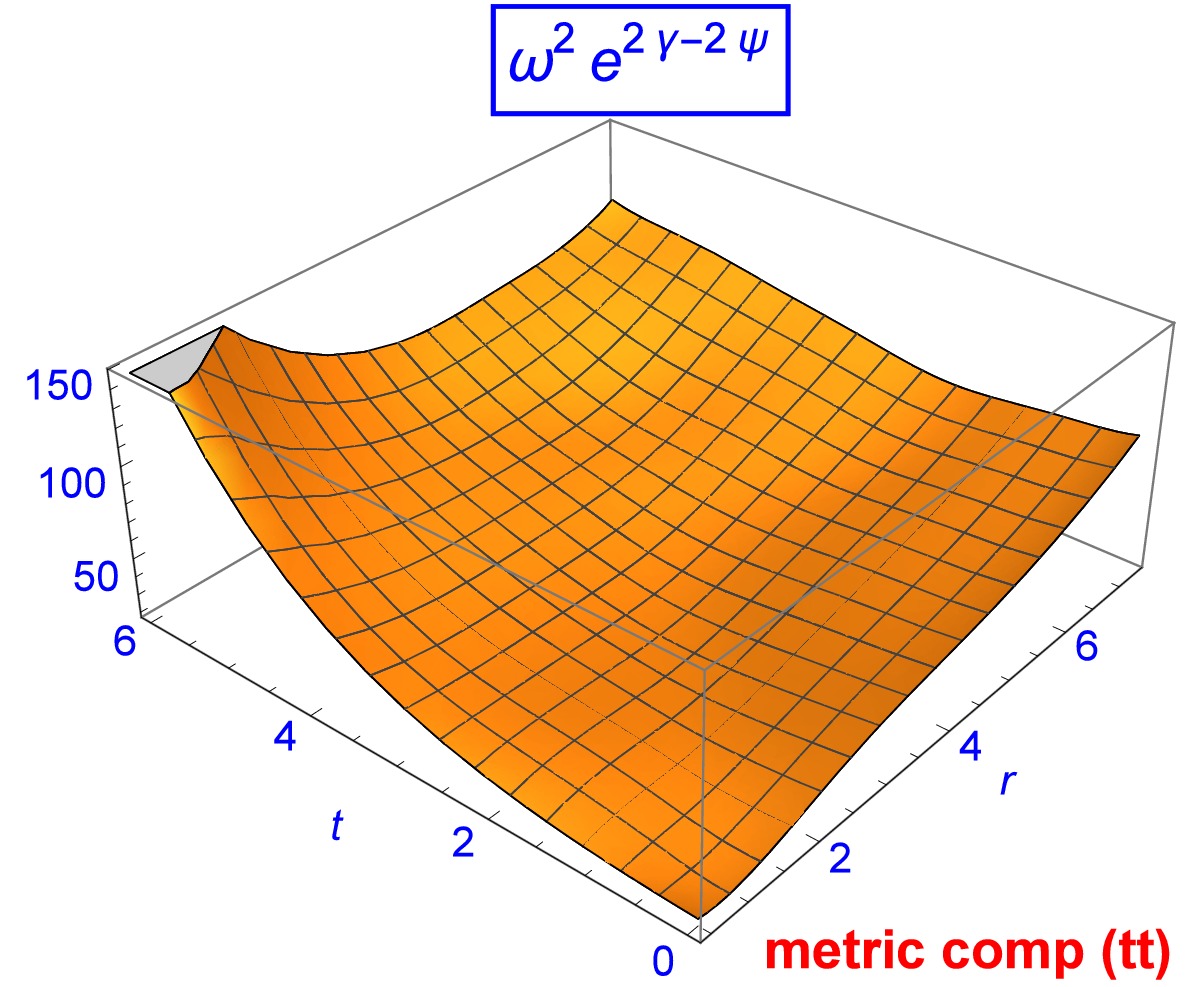}
\includegraphics[width=5cm]{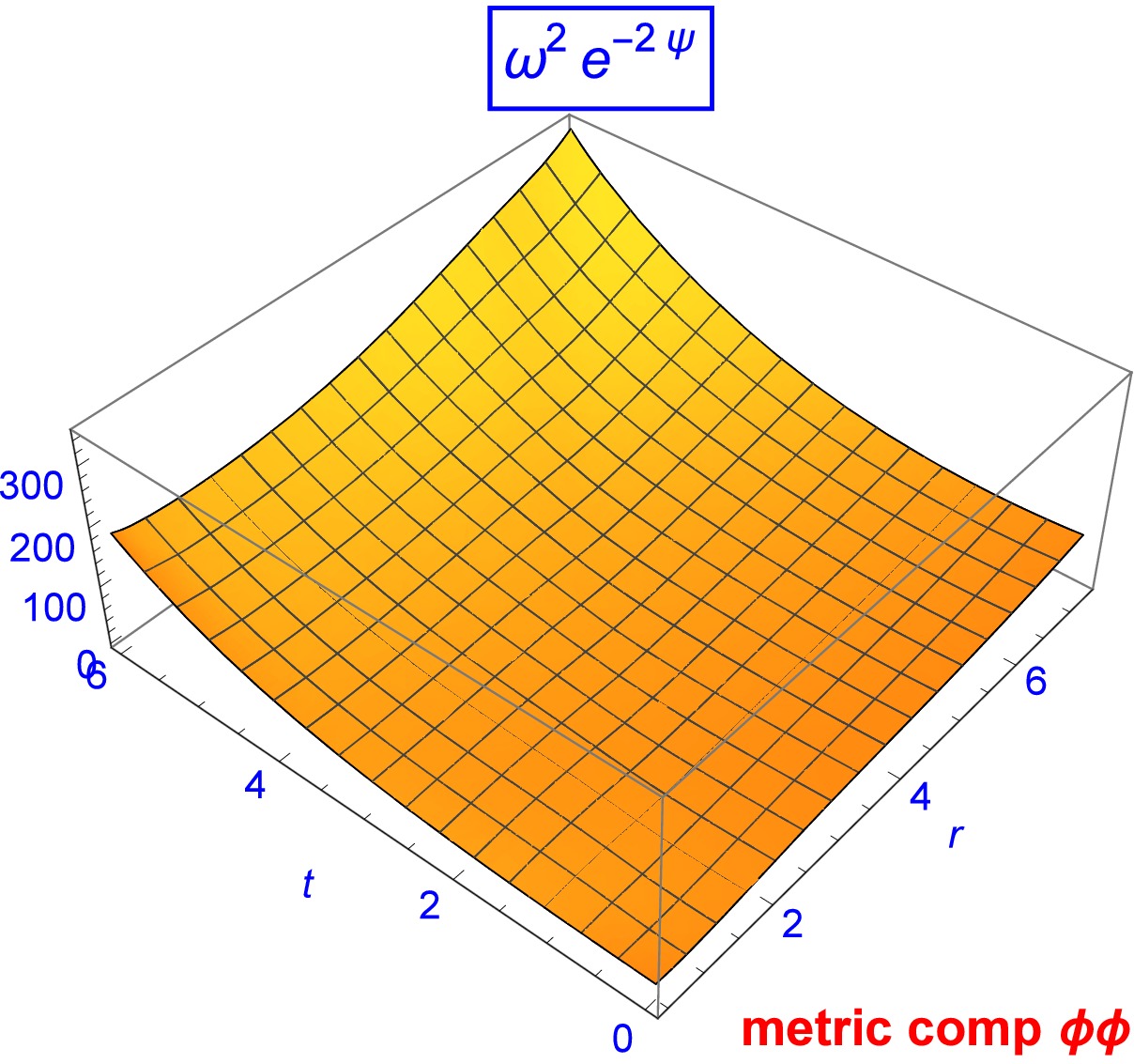}
\includegraphics[width=5cm]{xomega.jpg}}
\centerline{
\includegraphics[width=5cm]{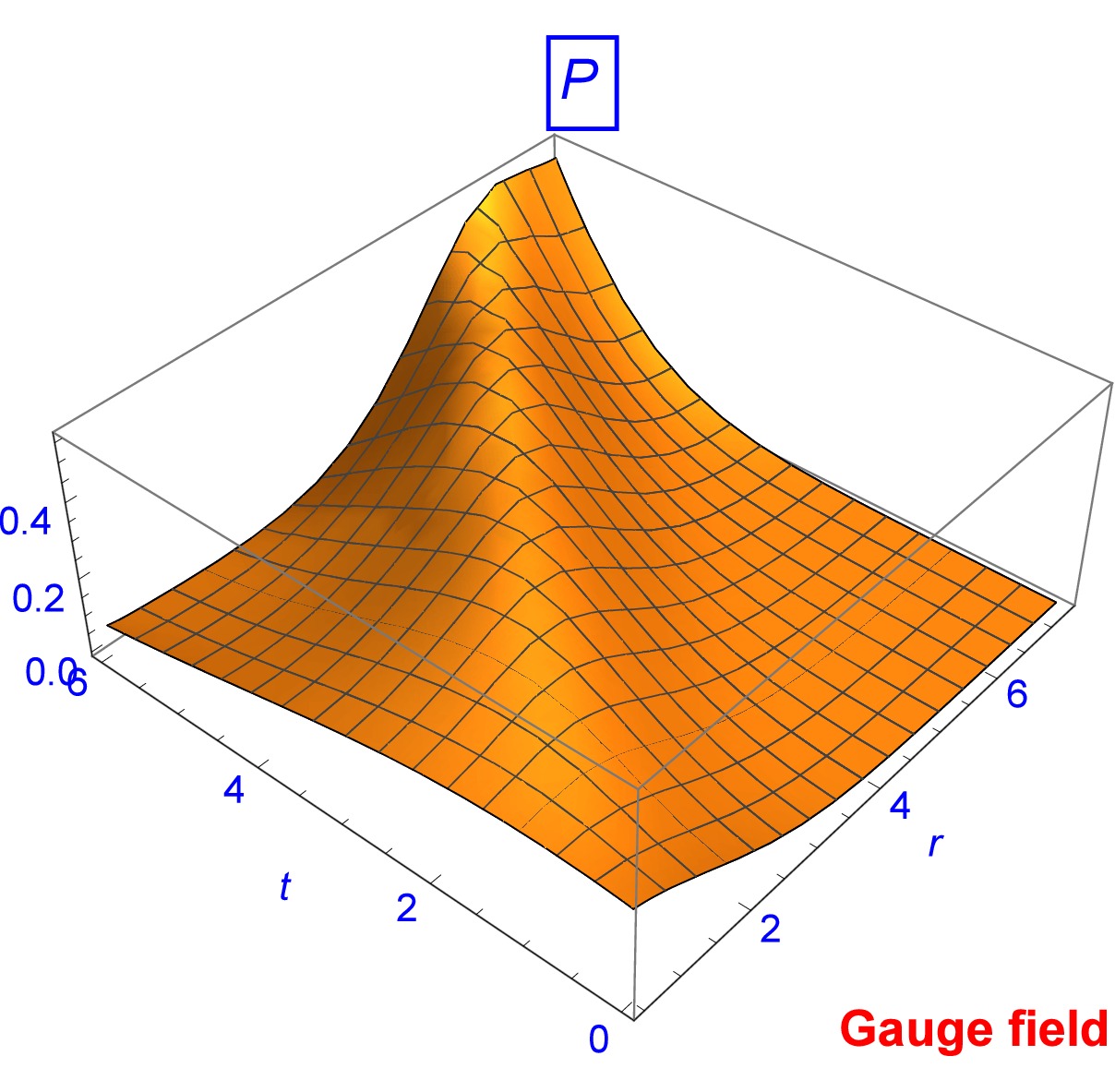}
\includegraphics[width=5cm]{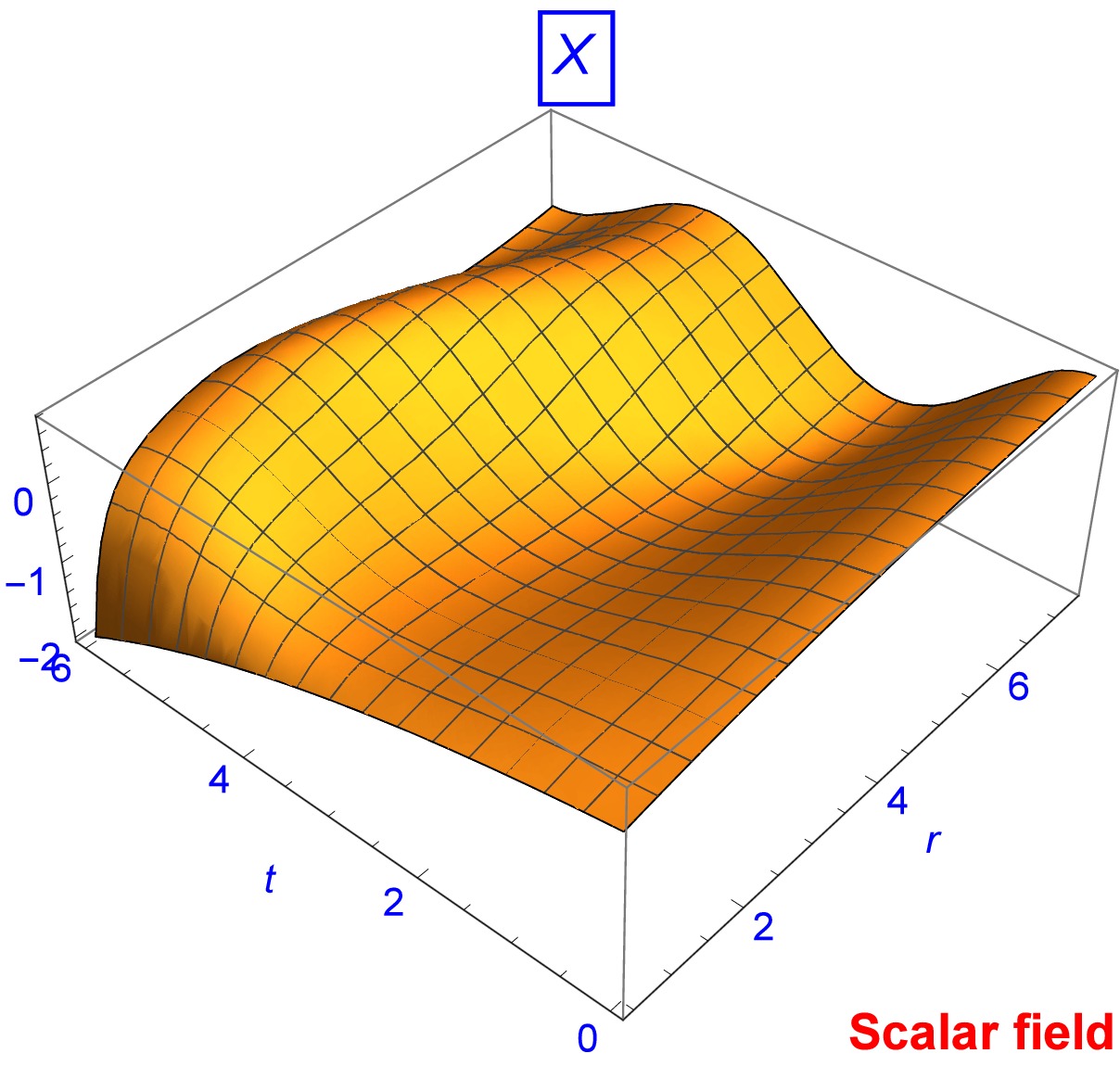}
\includegraphics[width=5cm]{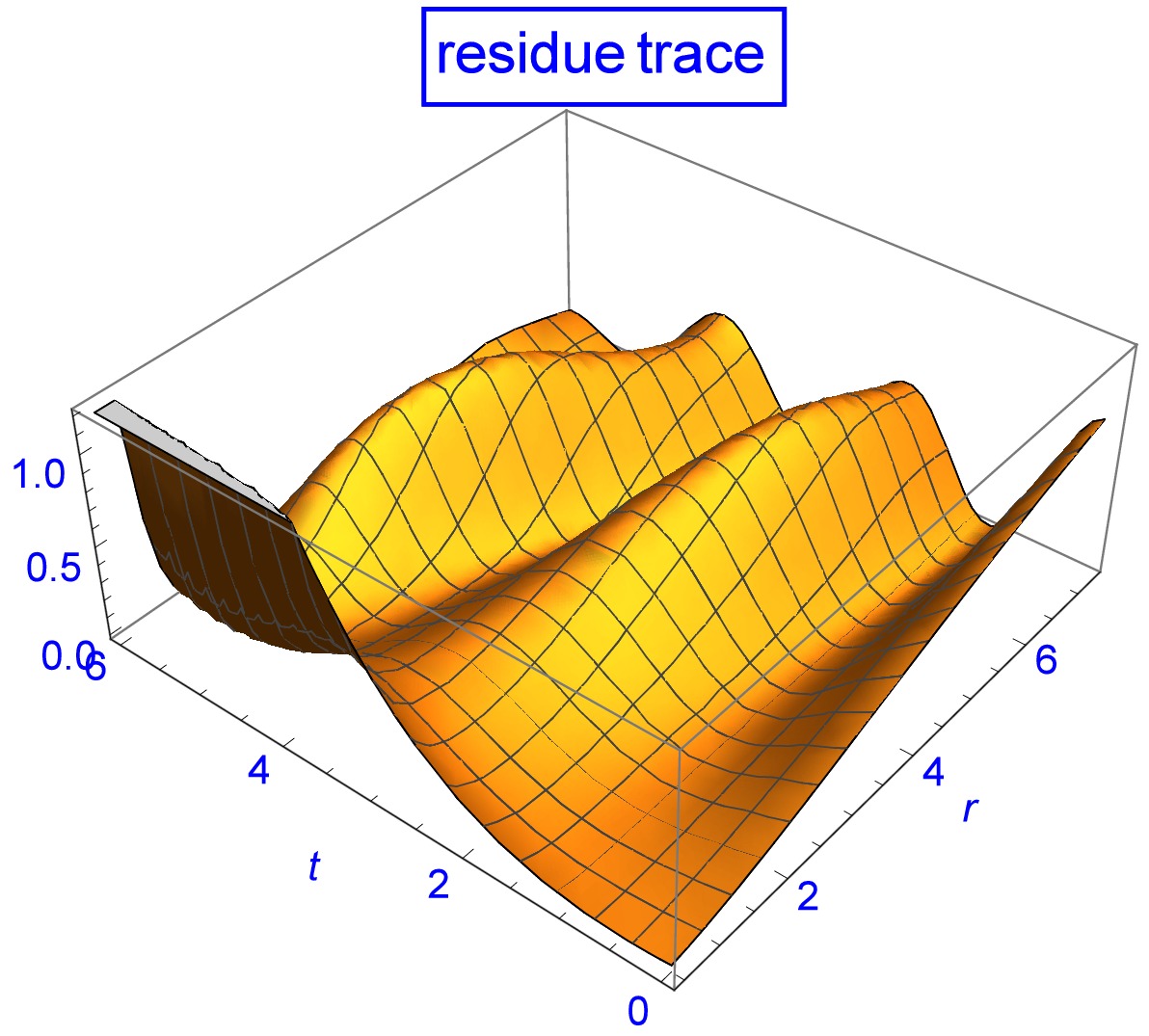}}
\vskip 0.4cm
\caption{A solution with oscillating behavior of the scalar field.}
\end{figure}

Our conjecture is that $\omega$, alias warp factor $W_1$, has a dual meaning. At a very early time in the evolution of our universe, when $\omega \rightarrow 0$, it describes the small-distance limit. At later times, it represents a warp (or scale) factor that determines the dynamical evolution of the universe.
\section{Conclusions}
\label{sec:9}
We analyzed conformal invariance in a non-vacuum 5D warped Einstein-scalar-gauge-field model, where the warp factor is reinterpreted as dilaton field. The equation for this dilaton can be isolated from the 5D field equations of the resulting un-physical metric.
On the "classical" level, the dilaton plays the role of a warp factor determining the evolution of the FLRW model.
On small scales, when the dilaton field approaches zero, no singular behavior emerges because the conformal invariant model treats the dilaton and Higgs field on equal footing.
The parameters of the scalar-gauge field, i.e., $(\beta, \eta, \epsilon)$ and the gravitational coupling constants enter the trace of the total energy momentum tensor constraint equation.
This means that the  mass of the vortex per unit length,  $\mu\approx 2\pi\eta^2\int g_{\varphi\varphi} dr$  and  the ratio of the gauge and scalar field, i. e. $\frac{\beta}{\epsilon^2}$, also determines the dilaton behavior.
The dimensionless parameter $\kappa_4^2\mu$ plays an important role in the physics of cosmic strings. It can be approximated by $ G\mu\approx\frac{\eta^2}{m_{pl}^2}$. Observational bounds require $G\mu\sim 10^{-6}$.
However, the most interesting gravitational impact will occur at $G\mu>>1$, which is possible in our 5D model by means of the warp factor, or now the dilaton field $\omega$.
If we specify $\omega$, then the evolution of $\omega^2\tilde g_{\mu\nu}$ becomes ambiguous: the time evolution of a FLRW model depends then heavily on the parameters determining the dilaton solution.
Conformal invariant gravity theories need additional constraint equations in order to obtain a traceless energy-momentum tensor.  If the conformal invariance is exact and spontaneously broken, then one needs additional field transformations on $\tilde g_{\mu\nu}$.
There are some shortcomings in our pure classical model.
First, one should like to incorporate fermionic fields. Secondly, all conformal anomalies of the model  must cancel out when one approaches smaller scales\cite{thooft4}. After all, we are dealing here with a curved $\tilde g_{\mu\nu}$. Some constraints must be fulfilled together with the "classical" tracelessness of the energy momentum tensor.
The reward is, however, that all the physical constants are in principal computable.


\begin{thebibliography}{30}
\bibitem{ran1}
Randall, L. and Sundrum, R. (1999) {\it Phys. Rev. Lett.}, {\bf 83},3370.
\bibitem{ran2}
Randall, L. and Sundrum, R. (1999) {\it Phys. Rev. Lett.} {\bf 83}, 4690.
\bibitem{roy1}
Maartens, R. (2007) {\it J. Phys. Conf. Ser.} {\bf 68}, 012046.
\bibitem{roy2}
Maartens, R. (2007)  {\it Lect. Notes. Phys.} {\bf 720}, 323.
\bibitem{roy3}
Maartens, R. and Koyama, K. (2010) {\it Living Rev. Relativity} {\bf 13}, 5
\bibitem{shir}
Shiromizu, T., Maeda, K. and Sasaki, M. (2000)  {\it Phys. Rev. D} {\bf 62}, 024012.
\bibitem{thooft5}
’t Hooft, G., (1993), gr-qc/9310026
\bibitem{thooft2}
’t Hooft, G., (2010), gr-qc/10090669v2
\bibitem{thooft3}
’t Hooft, G., (2011), {\it Found. of Phys.} {\bf 41}, 1829.
\bibitem{strom}
Strominger, A. (2001), {\it Journal of High Energy Physics.}  {\bf 1}), 34.
\bibitem{thooft1}
’t Hooft, G., (2010), gr-qc/10110061v1
\bibitem{thooft4}
’t Hooft, G., (2015), gr-qc/151104427v1
\bibitem{mann}
Mannheim, P. D., (2017), hep-th/161008907v2
\bibitem{wald}
Wald, R. M.  (1984)  General Relativity, University of Chicago Press, Chicago,  USA.
\bibitem{slag1}
Slagter, R.J. and Pan, S. (2016) {\it Found. of Phys.} {\bf 46}, 1075.
\bibitem{islam}
Islam, J. N., (1985) Rotating Fields in General Relativity, Cambridge University Press, Cambridge, UK.
\bibitem{perj}
Percacci, R. (2016)  An Introduction to Covariant Quantum Gravity and Asymptotic Safety, World Scientific, Singapore.
\bibitem{har}
Harada, T., Nakao, K. and Nolan, B. C. (2009) {\it Phys. Rev. D} {\bf 80}, 024025.
\bibitem{vand}
Garriga, J. and Vandaguer, E. (1987) {\it Phys. Rev. D} {\bf 36} 2250.
\bibitem{greg}
Gregory, R. (1989) {\it Phys. Rev. D} {\bf 39}, 2108.
\bibitem{and}
Anderson, M.R. (2003) The Mathematical Theory of Cosmic Strings. IoP publishing, Bistol, UK.
\bibitem{slag2}
Slagter, R.J. (2016) {\it J. of Mod.Phys.} {\bf 7}, 501.
\bibitem{slag3}
Slagter, R.J. (2017) {\it J. of Mod.Phys.} {\bf 8}, 163.
\bibitem{slag4}
Slagter, R.J. (2014) {\it Int. J. of Mod.Phys. D} {\bf 10}, 1237.
\bibitem{step}
Stephani, H., Kramer, D., Maccallum, M. and Herlt, E (2009) Exact Solutions of Einstein's Field Equations, Cambridge University Press, Cambridge, UK.
\bibitem{muk}
Mukhanov, V. F., Winitzki, S. (2007) Quantum Effects in Gravity, Cambridge University Press, Cambridge, UK.
\bibitem{park}
Parker, L. E. and Toms, D. J. (2009) Quantum Field Theory in Curved Spacetime, Cambridge University Press, Cambridge, UK.
\end{thebibliography}
\end{document}